\documentclass[pra,twocolumn,superscriptaddress,showpacs]{revtex4}

\usepackage{amsfonts}
\usepackage{amssymb}
\usepackage{amsmath}
\usepackage{hyperref}
\usepackage{graphicx}
\usepackage{longtable}

\newcommand{\calS}{\mathcal{S}}
\newcommand{\calA}{\mathcal{A}}
\newcommand{\psubpi}{p_{\pi}}
\newcommand{\ssb}{s_{b}}

\begin{document}
\title{Graph isomorphism and adiabatic quantum computing}

\date{\today}

\author{Frank Gaitan}
\affiliation{Laboratory for Physical Sciences, 8050 Greenmead Dr, 
College Park, MD 20740}

\author{Lane Clark}
\affiliation{Department of Mathematics, Southern Illinois University, 
Carbondale, IL 62901-4401}

\begin{abstract}
In the Graph Isomorphism (GI) problem two $N$-vertex graphs $G$ and 
$G^{\prime}$ are given and the task is to determine whether there exists 
a permutation of the vertices of $G$ that preserves adjacency and transforms 
$G\rightarrow G^{\prime}$. If yes, then $G$ and $G^{\prime}$ are said to be 
isomorphic; otherwise they are non-isomorphic. The GI problem is an important 
problem in computer science and is thought to be of comparable difficulty to
integer factorization. In this paper we present a quantum algorithm that solves
arbitrary instances of GI and which also provides a novel approach to
determining all automorphisms of a given graph. We show how the GI problem 
can be converted to a combinatorial optimization problem that can be solved 
using adiabatic quantum evolution. We numerically simulate the algorithm's 
quantum dynamics and show that it correctly: (i) distinguishes non-isomorphic 
graphs; (ii) recognizes isomorphic graphs and determines the permutation(s)
that connect them; and (iii) finds the automorphism 
group of a given graph $G$. We then discuss the GI quantum algorithm's 
experimental implementation, and close by showing how it can be leveraged 
to give a quantum algorithm that solves arbitrary instances of the 
NP-Complete Sub-Graph Isomorphism problem. The computational complexity
of an adiabatic quantum algorithm is largely determined by the minimum energy 
gap $\Delta (N)$ separating the ground- and first-excited states in the limit 
of large problem size $N\gg 1$. Calculating $\Delta (N)$ in this limit is a
fundamental open problem in adiabatic quantum computing, and so it is not
possible to determine the computational complexity of adiabatic quantum 
algorithms in general, nor consequently, of the specific adiabatic quantum 
algorithms presented here. Adiabatic quantum computing has been shown to be 
equivalent to the circuit-model of quantum computing, and so development of 
adiabatic quantum algorithms continues to be of great interest.
\end{abstract}


\maketitle

\section{Introduction}
\label{sec1}

An instance of the Graph Isomorphism (GI) problem is specified by two 
$N$-vertex graphs $G$ and $G^{\prime}$ and the challenge is to determine 
whether there exists a permutation of the vertices of $G$ that preserves 
adjacency and transforms $G\rightarrow G^{\prime}$. When such a permutation 
exists, the graphs are said to be isomorphic; otherwise they are non-isomorphic.
GI has been heavily studied in computer science \cite{kobler}. Polynomial 
classical algorithms exist for special cases of GI, still it has not been 
possible to prove that GI is in P. Although it is known that GI is in NP, it
has also not been possible to prove that it is NP-Complete. The situation is
the same for Integer Factorization (IF)---it belongs to NP, but is not known to be 
in P or to be NP-Complete. GI and IF are believed to be of comparable 
computational difficulty\cite{Aro&Bar}. 

IF and GI have also been examined from the perspective of quantum algorithms
and both have been connected to the hidden subgroup problem (HSP) \cite{jozsa}.
For IF the hidden subgroup is contained in an abelian parent group 
($Z^{\ast}_{n} =$ group of units modulo n), while for GI the parent group is 
non-abelian ($S_{n} =$ symmetric group on $n$ elements). While Fourier 
sampling allows the abelian HSP to be solved efficiently\cite{Bern&Vaz}, strong
Fourier sampling does not allow an efficient solution of the non-abelian HSP 
over $S_{n}$ \cite{moore}. At this time an efficient quantum 
algorithm for GI is not known.

A number of researchers have considered using the dynamics of physical systems 
to solve instances of GI. Starting from physically motivated conjectures, these 
approaches embed the structure of the graphs appearing in the GI instance
into the Hamiltonian that drives the system dynamics. In Ref.~\cite{Gud&Nuss}
the systems considered were classical, while
Refs.~\cite{Rudol,Shiau,Gamb,Rud,Hen} worked with quantum systems. 
\begin{itemize}
\item Building on Refs.~\cite{Rudol} and \cite{Shiau}, Refs.~\cite{Gamb} and
\cite{Rud} proposed using multi-particle quantum random walks (QRW) on graphs 
as a means for distinguishing pairs of non-isomorphic graphs. Numerical tests 
of this approach focused on GI instances involving strongly regular graphs 
(SRG). The adjacency matrix for each SRG was used to define the Hamiltonian 
$H(G)$ that drives the QRW on the graph $G$. The walkers can only hop between 
vertices joined by an edge in $G$. The propagator $U(G) = \exp [-iH(G)t]$ is 
evaluated at a fixed time $t$ for each SRG associated with a GI instance. The 
two propagators are used to define a comparison function that is conjectured to
vanish for isomorphic pairs of SRGs, and to be non-zero otherwise. 
Refs.~\cite{Gamb} and \cite{Rud} examined both interacting and non-interacting 
systems of quantum walkers and found that: (i)~no non-interacting QRW with a 
fixed number of walkers can distinguish all pairs of SRGs; (ii)~increasing the 
number of walkers increases the distinguishing power of the approach; and 
(iii)~two-interacting bosonic walkers have more distinguishing power than both
one and two non-interacting walkers. No analysis was provided of the 
algorithm's runtime $T(N)$ versus problem size $N$ in the limit of large 
problem size $N\gg 1$, and so the computational complexity of this GI 
algorithm is currently unknown. Note that for isomorphic pairs of graphs, 
this algorithm cannot determine the permutation(s) connecting the two 
graphs, nor the automorphism group of a given graph. Finally, no discussion 
of the algorithm's experimental implementation is given.

\item The GI algorithm presented in Ref.~\cite{Hen} is based on adiabatic
quantum evolution: here the problem Hamiltonian is an Ising Hamiltonian
that identifies the qubits with the vertices of a graph, and qubits $i$ and 
$j$ interact antiferromagnetically only when vertices $i$ and $j$ in the
graph are joined by an edge. It was conjectured that the instantaneous 
ground-state encodes enough information about the graph to allow suitably 
chosen measurements to distinguish pairs of non-isomorphic graphs. Physical 
observables that are invariant under qubit permutations are measured at 
intermediate times and the differences in the dynamics generated by two 
non-isomorphic graphs are assumed to allow the measurement outcomes to 
recognize the graphs as non-isomorphic. The algorithm was tested numerically 
on (mostly) SRGs, and it was found that combinations of measurements of 
the: (i)~spin-glass order parameter, (ii)~$x$-magnetization, and (iii)~total 
average energy allowed all the non-isomorphic pairs of graphs examined to be 
distinguished. Ref.~\cite{Hen} did not determine the scaling relation for the 
runtime $T(N)$ versus problem size $N$ for $N\gg 1$, and so the 
computational complexity of this algorithm is currently unknown.The
algorithm is also unable to determine the permutation(s) that connect 
a pair of isomorphic graphs, nor the automorphism group of a given graph. 
Ref.~\cite{Hen} discussed the experimental implementation of this algorithm 
and noted that the measurements available on the D-Wave hardware do not 
allow the observables used in the numerical tests to be measured on the 
hardware.
\end{itemize}

In this paper we present a quantum algorithm that solves arbitrary instances of
GI. The algorithm also provides a novel approach for determining the 
automorphism group of a given graph. The GI quantum algorithm is constructed 
by first converting an instance of GI into an instance of a combinatorial 
optimization problem whose cost function, by construction, has a zero minimum 
value when the pair of graphs in the GI instance are isomorphic, and is positive 
when the pair are non-isomorphic. The specification of the GI quantum algorithm
is completed by showing how the combinatorial optimization problem can be 
solved using adiabatic quantum evolution. To test the effectiveness of this GI 
quantum algorithm we numerically simulated its Schrodinger dynamics. The 
simulation results show that it can correctly: (i)~distinguish pairs of 
non-isomorphic graphs; (ii)~recognize pairs of isomorphic graphs and determine
the permutation(s) that connect them; and (iii)~find the automorphism group 
of a given graph. We also discuss the experimental implementation of the GI 
algorithm, and show how it can be leveraged to give a quantum algorithm that 
solves arbitrary instances of the (NP-Complete) Sub-Graph Isomorphism (SGI) 
problem. As explained in Section~\ref{sec4}, calculation of the runtime for an 
adiabatic quantum algorithm in the limit of large problem size is a fundamental 
open problem in adiabatic quantum computing. It is thus not presently possible 
to determine the computational complexity of adiabatic quantum algorithms in 
general, nor, consequently, of the specific adiabatic quantum algorithms 
presented here. However, because adiabatic quantum computing has been shown 
to be equivalent to the circuit-model of quantum computing 
\cite{ahar,kempe,terhal}, the development of adiabatic quantum algorithms 
continues to be of great interest. Just as with the GI algorithms of 
Refs.~\cite{Shiau}-\cite{Hen}, our GI algorithm also has unknown complexity.
However, unlike the algorithms of Refs.~\cite{Shiau}-\cite{Hen}, the GI 
algorithm presented here: (i)~encodes the GI instance explicitly into the cost 
function of a combinatorial optimization problem which is solved using 
adiabatic quantum evolution without introducing physical conjectures; and 
(ii)~determines the permutation(s) connecting two isomorphic graphs, and the 
automorphism group of a given graph. As we shall see, our GI algorithm can 
be implemented on existing D-Wave hardware using established embedding 
procedures \cite{bian}. Such an implementation is in the works and will be 
reported elsewhere.

The structure of this paper is as follows. In Section~\ref{sec2} we give a 
careful presentation of the GI problem, and show how an instance of GI 
can be converted to an instance of a combinatorial optimization problem whose
solution  (Section~\ref{sec3}) can be found using adiabatic quantum evolution. 
To test the performance of the GI quantum algorithm introduced in 
Section~\ref{sec3}, we numerically simulated its Schrodinger dynamics and the 
results of that simulation are presented in Section~\ref{sec4}. In 
Section~\ref{sec5} we describe the experimental implementation of the GI 
algorithm, and in Section~\ref{sec6} we show how it can be used to give a 
quantum algorithm that solves arbitrary instances of the NP-Complete problem 
known as SubGraph Isomorphism. Finally, we summarize our results in 
Section~\ref{sec7}. Two appendices are also included. The first briefly
summarizes the quantum adiabatic theorem and its use in adiabatic quantum 
computing, and the second reviews the approach to embedding the problem 
Hamiltonian for an adiabatic quantum algorithm onto the D-Wave hardware
that was presented in Ref.~\cite{bian}.

\section{Graph isomorphism problem}
\label{sec2}

In this Section we introduce the Graph Isomorphism (GI) problem and show how 
an instance of GI can be converted into an instance of a combinatorial 
optimization problem (COP) whose cost function has zero (non-zero) minimum
value when the pair of graphs being studied are isomorphic (non-isomorphic).

\subsection{Graphs and graph isomorphism}
\label{sec2a}

A graph $G$ is specified by a set of vertices $V$ and a set of edges $E$.  We 
focus on \textit{simple\/} graphs in which an edge only connects distinct 
vertices, and the edges are undirected. The \textit{order of} $G$ is defined 
to be the number of vertices contained in $V$, and two vertices are said to be
\textit{adjacent\/} if they are connected by an edge. If $x$ and $y$ are 
adjacent, we say that $y$ is a \textit{neighbor\/} of $x$, and vice versa. 
The \textit{degree\/} $d(x)$ of a vertex $x$ is equal to the number of vertices 
that are adjacent to $x$. The \textit{degree sequence\/} of a graph lists the 
degree of each vertex in the graph ordered from largest degree to smallest. 
A graph $G$ of order $N$ can also be specified by its 
\textit{adjacency matrix\/} $A$ which is an $N\times N$ matrix whose matrix 
element $a_{i,j}= 1\; (0)$ if the vertices $i$ and $j$ are (are not) adjacent. 
For simple graphs $a_{i,i}=0$ and $a_{i,j}= a_{j,i}$ since edges only connect 
distinct vertices and are undirected.

Two graphs $G$ and $G^{\prime}$ are said to be \textit{isomorphic\/} if there is
a one-to-one correspondence $\pi$ between the vertex sets $V$ and $V^{\prime}$
such that two vertices $x$ and $y$ are adjacent in $G$ if and only if their 
images $\pi_x$ and $\pi_y$ are adjacent in $G^{\prime}$. The graphs $G$ and 
$G^{\prime}$ are non-isomorphic if no such $\pi$ exists. Since no one-to-one 
correspondence $\pi$ can exist when the number of vertices in $G$ and 
$G^{\prime}$ are different, graphs with unequal orders are always 
non-isomorphic. It can also be shown \cite{Chart&Zhang} that if two graphs 
are isomorphic, they must have identical degree sequences.

We can also describe graph isomorphism in terms of the adjacency matrices $A$
and $A^{\prime}$ of the graphs $G$ and $G^{\prime}$, respectively. The graphs
are isomorphic if and only if there exists a permutation matrix $\sigma$ of the
vertices of $G$ that satisfies
\begin{equation}
A^{\prime} = \sigma A\sigma^{T} ,
\label{GIcond}
\end{equation}
where $\sigma^{T}$ is the transpose of $\sigma$. It is straight-forward to show 
that if $G$ and $G^{\prime}$ are isomorphic, then a permutation matrix $\sigma$
exists that satisfies Eq.~(\ref{GIcond}). To prove the ``only if'' statement, 
note that the $i$-$j$ matrix element of the RHS is $A_{\sigma_i,\sigma_j}$. 
Eq.~(\ref{GIcond}) is thus satisfied when a permutation matrix $\sigma$ exists 
such that $A^{\prime}_{i,j} = A_{\sigma_i,\sigma_j}$. Thus Eq.~(\ref{GIcond}) 
is simply the condition that $\sigma$ preserve adjacency. Since a permutation 
is a one-to-one correpondence, the existence of a permutation matrix $\sigma$ 
satisfying Eq.~(\ref{GIcond}) implies $G$ and $G^{\prime}$ are isomorphic.  
Thus, the existence of a permutation matrix $\sigma$ satisfying 
Eq.~(\ref{GIcond}) is an equivalent way to define graph isomorphism. 

The Graph Isomorphism (GI) problem is to determine whether two given graphs
$G$ and $G^{\prime}$ are isomorphic. The problem is only non-trivial when $G$
and $G^{\prime}$ have the same order and so we focus on that case in this paper.

\subsection{Permutations, binary strings, and linear maps}
\label{sec2b}

A permutation $\pi$ of a finite set $\mathcal{S} =   \{ 0,\ldots , N-1 \}$ is
a one-to-one correspondence from $\calS \rightarrow\calS$ which sends $i
\rightarrow \pi_{i}$ such that $\pi_{i} \in\calS$, and $\pi_{i}\neq 
\pi_{j}$ for $i\neq j$. The permutation $\pi$ can be written 
\begin{equation}
\pi = \left(  \begin{array}{ccccc}
                       0 & \cdots & i & \cdots & N-1\\
                      \pi_{0} & \cdots & \pi_{i} & \cdots & \pi_{N-1}
                  \end{array}
        \right) ,
\label{permdef}
\end{equation}
where column $i$ indicates that $\pi$ sends $i\rightarrow \pi_{i}$. Since the
top row on the RHS of Eq.~(\ref{permdef}) is the same for all permutations,
all the information about $\pi$ is contained in the bottom row. Thus we can map
a permutation $\pi$ into an integer string $P(\pi ) = \pi_{0}\cdots \pi_{N-1}$, 
with $\pi_{i}\in\calS$ and $\pi_{i}\neq \pi_{j}$ for $i\neq j$.

For reasons that will become clear in Section~\ref{sec2c}, we want to convert 
the integer string $P(\pi ) = \pi_{0}\cdots \pi_{N-1}$ into a binary string 
$\psubpi$. This can be done by replacing each $\pi_{i}$ in $P(\pi )$ by the 
unique binary string formed from the coefficients appearing in its binary 
decomposition
\begin{equation}
\pi_{i} = \sum_{j=0}^{U-1} \pi_{i,j} \left( 2\right)^{j} .
\label{bindecomp}
\end{equation}
Here $U \equiv \lceil \log_{2} N \rceil$. Thus the integer string $P(\pi )$ is 
transformed to the binary string
\begin{equation}
\psubpi = \left(\pi_{0,0}\cdots \pi_{0,U-1}\right) \cdots \left( \pi_{N-1,0} 
               \cdots         \pi_{N-1,U-1}\right),
\label{permbinstr}
\end{equation}
where $\pi_{i,j}\in\{ 0,1\}$. The binary string $\psubpi$ has length $NU$, where
\begin{equation}
N \leq 2^{U} \equiv M+1.
\label{Mdef}
\end{equation}
Thus we can identify a permutation $\pi$ with the binary string $\psubpi$ in 
Eq.~(\ref{permbinstr}).

Let $\mathcal{H}$ be the Hamming space of binary strings of length $NU$. This 
space contains $2^{NU}$ strings, and we have just seen that $N!$ of these 
strings $\psubpi$ encode permutations $\pi$. Our last task is to define a 
mapping from $\mathcal{H}$ to the space of $N\times N$ matrices $\sigma$ with 
binary matrix elements $\sigma_{i,j} = 0,1$. The mapping is constructed as 
follows:
\begin{enumerate}
\item Let $s_{b} = s_{0}\cdots s_{NU -1}$ be a binary string in $\mathcal{H}$.
We parse $s_{b}$ into $N$ substrings of length U as follows:
\begin{equation}
s_{b} =\left(s_{0}\cdots s_{U-1}\right)\,\left( s_{U}\cdots
                s_{2U-1}\right) \cdots\left( s_{(N-1)U} \cdots
                  s_{NU-1}\right).
\label{parsestr}
\end{equation}
\item For each substring $s_{iU}\cdots s_{(i+1)U - 1}$, construct the integer
\begin{equation}
s_{i} = \sum_{j=0}^{U-1} s_{iU+j}\left( 2\right)^{j} \leq 2^{U} - 1 = M.
\label{substr2int}
\end{equation}
\item Finally, introduce the integer string $s = s_{0}\cdots s_{N-1}$, and 
define the $N\times N$ matrix $\sigma (s)$ to have matrix elements
\begin{equation}
\sigma_{i,j}(s) = 
\left\{ \begin{array}{cl}
               0,\hspace{0.2in} {} & \mathrm{if}\: s_{j} > N-1\\
               \delta_{i,s_{j}}, & \mathrm{if} \: 0 \leq s_{j} \leq N-1 ,
           \end{array} \right.
\label{sigmatelm}
\end{equation}
where $i,j\in\calS$, and $\delta_{x,y}$ is the Kronecker delta.
\end{enumerate}
Note that when the binary string $s_{b}$ corresponds to a permutation, the 
matrix $\sigma (s)$ is a permutation matrix since the $s_{i}$ formed in 
step~$2$ will obey $0\leq s_{i}\leq N-1$ and $s_{i}\neq s_{j}$ for $i\neq j$. 
In this case, if $A$ is the adjacency matrix for a graph $G$, then $A^{\prime} =
\sigma (s) A \sigma^{T}(s)$ will be the adjacency matrix for a graph 
$G^{\prime}$ isomorphic to $G$. On the other hand, if $s_{b}$ does not 
correspond to a permutation, then the adjacency matrix $A^{\prime} = 
\sigma (s)A\sigma^{T}(s)$ must correspond to a graph $G^{\prime}$ which 
is not isomorphic to $G$.

The result of our development so far is the establishment of a map from binary
strings of length $NU$ to $N\times N$ matrices (viz.~linear maps) with binary
matrix elements. When the string is (is not) a permutation, the matrix
produced is (is not) a permutation matrix. Finally, recall from 
Stirling's formula that $\log_{2} N! \sim N \log_{2} N -  N$ which is the 
number of bits needed to represent $N!$. Our encoding of permutations 
uses $N \lceil \log_{2} N\rceil$ bits and so approaches asymptotically what 
is required by Stirling's formula. 

\subsection{Graph isomorphism and combinatorial optimization}
\label{sec2c}

As seen above, an instance of GI is specified by a pair of graphs $G$ and
$G^{\prime}$ (or equivalently, by a pair of adjacency matrices $A$ and
$A^{\prime}$). Here we show how a GI instance can be transformed into an
instance of a combinatorial optimization problem (COP) whose cost function has a
minimum value of zero if and only if $G$ and $G^{\prime}$ are isomorphic.

The search space for the COP is the Hamming space $\mathcal{H}$ of binary 
strings $s_{b}$ of length $NU$ which are associated with the integer strings 
$s$ and matrices $\sigma (s)$ introduced in Section~\ref{sec2b}. The COP cost 
function $C(s)$ contains three contributions
\begin{equation}
C(s) = C_{1}(s) + C_{2}(s) + C_{3}(s) .
\label{costfuncdef}
\end{equation}
The first two terms on the RHS penalize integer strings $s = s_{0}\cdots 
s_{N-1}$ whose associated matrix $\sigma (s)$ is not a permutation matrix,
\begin{eqnarray}
C_{1}(s) & = & \sum_{i=0}^{N-1}\sum_{\alpha = N}^{M} \delta_{s_{i},\alpha}
      \label{C1defv0} \\
C_{2}(s) & = & \sum_{i=0}^{N-2}\sum_{j = i+1}^{N-1} \delta_{s_{i},s_{j}} ,
     \label{C2defv0}
\end{eqnarray}
where $\delta_{x,y}$ is the Kronecker delta. 
We see that $C_{1}(s) > 0$ when $s_{i} > N - 1$ for some $i$, and $C_{2}(s)
> 0$ when $s_{i}=s_{j}$ for some $i\neq j$. Thus $C_{1}(s) + C_{2}(s) = 0$ if 
and only if $\sigma (s)$ is a permutation matrix. The third term $C_{3}(s)$ adds
a penalty when $\sigma (s)A\sigma^{T}(s) \neq A^{\prime}$:
\begin{equation}
C_{3}(s) = \| \sigma (s)A\sigma^{T}(s) - A^{\prime}\|_{i} .
   \label{C3defv0}
\end{equation}
Here $\| M\|_{i}$ is the $L_{i}$-norm of $M$. In the numerical simulations 
discussed in Section~\ref{sec4}, the $L_{1}$-norm is used, though any 
$L_{i}$-norm would be acceptable. Thus, when $G$ and 
$G^{\prime}$ are isomorphic, $C_{3}(s) = 0$, and $\sigma (s)$ is the permutation
of vertices of $G$ that maps $G\rightarrow G^{\prime}$. Putting all these 
remarks together, we see that if $C(s) = 0$ for some integer string $s$, then 
$G$ and $G^{\prime}$ are isomorphic and $\sigma (s)$ is the permutation that 
connects them. On the other hand, if $C(s) > 0$ for all strings $s$, then $G$ 
and $G^{\prime}$ are non-isomorphic.

We have thus converted an instance of GI into an instance of the following COP:
\begin{description}
\item[Graph Isomorphism COP:] Given the $N$-vertex graphs $G$ and $G^{\prime}$
and the associated cost function $C(s)$ defined above, find an integer string 
$s_{\ast}$ that minimizes $C(s)$.
\end{description}
By construction: (i)~$C(s_{\ast}) = 0$ if and only if $G$ and $G^{\prime}$ are
isomorphic and $\sigma (s_{\ast})$ is the permutation matrix mapping 
$G\rightarrow G^{\prime}$; and (ii)~$C(s_{\ast}) > 0$ if and only if $G$ and 
$G^{\prime}$ are non-isomorphic. 

Before moving on, notice that if $G = G^{\prime}$, then $C(s_{\ast}) = 0$ 
since $G$ is certainly isomorphic to itself. In this case $\sigma (s_{\ast})$ 
is an automorphism of $G$. We shall see that the GI quantum algorithm to be 
introduced in Section~\ref{sec3} provides a novel approach for finding the 
automorphism group of a graph. 

\section{Adiabatic quantum algorithm for graph isomorphism}
\label{sec3}

A quantum algorithm is an algorithm that can be run on a 
realistic model of quantum computation \cite{mosca}. One such model is 
adiabatic quantum computation \cite{aharonov} which is based on adiabatic 
quantum evolution \cite{messiah,Schiff}. The adiabatic quantum optimization 
(AQO) algorithm \cite{farhi} is an example of adiabatic quantum computation 
that exploits the adiabatic dynamics of a quantum system to solve a COP (see
Appendix~\ref{appendixQAT} for a brief overview). The AQO 
algorithm uses the optimization problem cost function to define a problem 
Hamiltonian $H_{P}$ whose ground-state subspace encodes all problem solutions. 
The algorithm evolves the state of an $L$-qubit register from the ground-state 
of an initial Hamiltonian $H_{i}$ to the ground-state of $H_{P}$ with 
probability approaching $1$ in the adiabatic limit. An appropriate measurement 
at the end of the adiabatic evolution yields a solution of the optimization 
problem almost certainly. The time-dependent Hamiltonian $H(t)$ for global AQO 
is
\begin{equation}
H(t) = \left( 1 - \frac{t}{T}\right) H_{i} + \left(\frac{t}{T}\right) H_{P},
\label{TDHam}
\end{equation}
where $T$ is the algorithm runtime, and adiabatic dynamics corresponds to
$T\rightarrow\infty$.

To map the GI COP onto an adiabatic quantum computation, we begin by promoting
the binary strings $s_{b}$ to computational basis states (CBS) $|s_{b}\rangle$.
Thus each bit in $s_{b}$ is promoted to a qubit so that the quantum register 
contains $L = NU = N\lceil \log_{2} N\rceil$ qubits. The CBS are defined to be
the $2^{L}$ eigenstates of $\sigma_{z}^{0}\otimes \cdots\otimes
\sigma_{z}^{L-1}$. The problem Hamiltonian $H_{P}$ is defined to be diagonal 
in the CBS with eigenvalue $C(s)$, where $s$ is the integer string associated 
with $\ssb$:
\begin{equation}
H_{P}|\ssb \rangle = C(s)|\ssb \rangle .
\label{HPdef}
\end{equation}
Note that (see Section~\ref{sec2c}) the ground-state energy of $H_{P}$ will be 
zero if and only if the graphs $G$ and $G^{\prime}$ are isomorphic. We will 
discuss the experimental realization of $H_{P}$ in Section~\ref{sec5}. The 
initial Hamiltonian $H_{i}$ is chosen to be
\begin{equation}
H_{i} = \sum_{l=0}^{L-1} \frac{1}{2}\left( I^{l} - \sigma_{x}^{l}\right) ,
\label{HIdef}
\end{equation}
where $I^{l}$ and $\sigma_{x}^{l}$ are the identity and x-Pauli operator for
qubit $l$, respectively. The ground-state of $H_{i}$ is the easily constructed
uniform superposition of CBS.

 The quantum algorithm for GI begins by preparing the $L$ qubit register in 
the ground-state of $H_{i}$ and then driving the qubit register dynamics using
the time-dependent Hamiltonian $H(t)$. At the end of the evolution the qubits
are measured in the computational basis. The outcome is the bit string 
$\ssb^{\ast}$ so that the final state of the register is $|\ssb^{\ast}\rangle$
and its energy is $C(s^{\ast})$, where $s^{\ast}$ is the integer string 
derived from $\ssb^{\ast}$. In the adiabatic limit, $C(s^{\ast})$ will be 
the ground-state energy, and if $C(s^{\ast}) =0$ ($\: > 0$) the algorithm
decides $G$ and $G^{\prime}$ are isomorphic (non-isomorphic). Note that any real
application of AQO will only be approximately adiabatic. Thus the probability 
that the final energy $C(s^{\ast})$ will be the ground-state energy will be
$1-\epsilon$. In this case the GI quantum algorithm must be run $k\sim 
\mathcal{O}(\ln (1-\delta )/\ln\epsilon )$ times so that, with probability 
$\delta > 1-\epsilon$, at least one of the measurements will return the 
ground-state energy. We can make $\delta$ arbitrarily close to $1$ by 
choosing $k$ sufficiently large.

\section{Numerical simulation of adiabatic quantum algorithm}
\label{sec4}

In this Section we present the results of a numerical simulation of the 
dynamics of the GI adiabatic quantum algorithm (AQA). Because the GI AQA uses 
$N\lceil \log_{2} N\rceil$ qubits, and these simulations were carried out using 
a classical digital computer, we were limited to GI instances involving graphs 
with order $N\leq 7$. Although we would like to have examined larger graphs, 
this simply was not practical. Note that the $N=7$ simulations use $21$ qubits. 
These simulations are at the upper limit of $20$-$22$ qubits at which simulation
of the full adiabatic Schrodinger dynamics is feasible 
\cite{Farhi2,Gaitan1,Gaitan2}. To simulate a GI instance with graphs of order 
$N=8$ requires a $24$ qubit simulation which is well beyond what can be done 
practically. The protocol for the simulations presented here follows 
Refs.~\cite{Farhi2,Gaitan1,Gaitan2,Gaitan3}. 

As explained in Appendix~\ref{appendixQAT}, the runtime for an adiabatic 
quantum algorithm is related to the minimum energy gap arising during the 
course of the adiabatic quantum evolution. Thus, determining the runtime 
scaling relation $T(N)$ versus problem size $N$ in the asymptotic limit 
($N \gg1$), reduces to determining the minimum gap scaling relation 
$\Delta (N)$ for large $N$. This, however, is a well-known, fundamental open 
problem in adiabatic quantum computing, and so it was not possible to determine
the asymptotic runtime scaling relation for our GI AQA. Although our 
numerical simulations could be used to compute a runtime for each of the GI 
instances considered below, we did not do so for two reasons. First, as noted 
above, the GI instances that can be simulated using a digital computer are 
limited to graphs with no more than seven vertices. These instances are thus 
far from the large problem-size limit $N\gg 1$, and so the associated runtimes 
tell us nothing about the asymptotic performance of the GI AQA. Second, it is 
well-known that the minimum energy gap encountered during adiabatic quantum 
evolution (and which largely determines the runtime) is sensitive to the 
particular Hamiltonian path followed by the adiabatic quantum evolution 
\cite{Farhi3,Rol&Cerf}. Determining the optimal Hamiltonian 
path which yields the largest minimum gap, and thus the shortest possible 
runtime, is another fundamental open problem in adiabatic quantum computing. 
As a result, the Hamiltonian path used in the numerical simulations (i.~e.\ the 
linear interpolating Hamiltonian $H(t)$ in Eq.~(\ref{TDHam})) will almost 
certainly be non-optimal, and so the runtime it produces will also, almost 
certainly, be non-optimal, and thus a poor indicator of GI AQA performance.

In Section~\ref{sec4a} we present simulation results for simple examples of 
isomorphic and non-isomorphic graphs. These examples allow us to illustrate 
the analysis of the simulation results in a simple setting. Section~\ref{sec4b} 
then presents our simulation results for non-isomorphic instances of: 
(i)~iso-spectral graphs; and (ii)~strongly regular graphs. Finally,
Section~\ref{sec4c} considers GI instances where $G^{\prime} = G$. Clearly,
all such instances correspond to isomorphic graphs since the identity permutation
will always map $G\rightarrow G$ and preserve adjacency. The situation is more
interesting when $G$ has symmetries which allow non-trivial permutations
as graph isomorphisms. These self-isomorphisms are referred to as graph 
automorphisms, and they form a group known as the automorphism group 
$Aut(G)$ of $G$. In this final subsection we use the GI AQA to find $Aut(G)$
for a number of graphs. To the best of our knowledge, using a GI algorithm to
find $Aut(G)$ is new. In all GI instances considered in this 
Section, the GI AQA correctly: (i)~distinguished non-isomorphic pairs of graphs; 
(ii)~recognized isomorphic pairs of graphs; and (iii)~determined the 
automorphism group of a given graph.

\subsection{Illustrative examples}
\label{sec4a}

Here we present the results of a numerical simulation of the GI AQA applied to 
two simple GI instances. In Section~\ref{sec4a1} (Section~\ref{sec4a2}) we 
examine an instance of two non-isomorphic (isomorphic) graphs. For the 
isomorphic instance we also present the permutations found by the GI AQA that
transforms $G$ into $G^{\prime}$ while preserving adjacency.

\subsubsection{Non-isomorphic graphs}
\label{sec4a1}

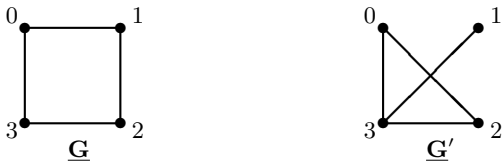
\begin{figure}[h!]
\begin{center}
\setlength{\unitlength}{0.05in}
\begin{picture}(90,15)(0,-1)
\thicklines
\put(10,12.5){\circle*{1}}
\put(20,12.5){\circle*{1}}
\put(10,2.5){\circle*{1}}
\put(20,2.5){\circle*{1}}
\put(10,2.5){\line(0,1){10}}
\put(20,2.5){\line(0,1){10}}
\put(10,2.5){\line(1,0){10}}
\put(10,12.5){\line(1,0){10}}
\put(8,13){$0$}
\put(8,1.0){$3$}
\put(21.2,13){$1$}
\put(21.2,1.0){$2$}
\put(14.5,-1){\textbf{\underline{G}}}

\put(47.5,12.5){\circle*{1}}
\put(57.5,12.5){\circle*{1}}
\put(47.5,2.5){\circle*{1}}
\put(57.5,2.5){\circle*{1}}
\put(45.5,13){$0$}
\put(45.5,1.0){$3$}
\put(58.7,13){$1$}
\put(58.7,1.0){$2$}
\put(47.5,2.5){\line(0,1){10}}
\put(47.5,2.5){\line(1,0){10}}
\put(47.5,12.5){\line(1,-1){10}}
\put(47.5,2.5){\line(1,1){10}}
\put(52,-1){$\mathbf{\underline{G}^{\prime}}$}
\end{picture}
\end{center}
\caption{\label{fig1}Two non-isomorphic $4$-vertex graphs $G$ and $G^{\prime}$.}
\end{figure}

Here we use the GI AQA to examine a GI instance in which the two graphs $G$
and $G^{\prime}$ are non-isomorphic. The two graphs are shown in 
Figure~\ref{fig1}.
Each graph contains $4$ vertices and $4$ edges, however they are 
non-isomorphic. Examining Figure~\ref{fig1} we see that the degree sequence
for $G$ is $\{ 2,2,2,2\}$, while that for $G^{\prime}$ is $\{ 3,2,2,1\}$. Since
these degree sequences are different we know that $G$ and $G^{\prime}$
are non-isomorphic. Finally, the adjacency matrices $A$ and $A^{\prime}$ for
$G$ and $G^{\prime}$, respectively, are:
\begin{equation}
A = \left(
                \begin{array}{cccc}
                   0 & 1 & 0 & 1 \\
                   1 & 0 & 1 & 0 \\
                   0 & 1 & 0 & 1 \\
                   1 & 0 & 1 & 0 \\
               \end{array}
      \right) 
\hspace{0.15in} ; \hspace{0.15in}  
A^{\prime} = \left(
                               \begin{array}{cccc}
                                  0 & 0 & 1 & 1 \\
                                  0 & 0 & 0 & 1 \\
                                  1 & 0 & 0 & 1 \\
                                  1 & 1 & 1 & 0 \\
                               \end{array}
                       \right) .
\label{twoadmat1}
\end{equation}

The GI AQA finds a non-zero final ground-state energy $E_{gs} = 4$, and so
correctly identifies these two graphs as non-isomorphic. It also finds that the 
final ground-state subspace has a degeneracy of $16$. The linear maps associated
with the $16$ CBS that span this subspace give rise to the lowest cost 
linear maps of $G$ relative to $G^{\prime}$. As these graphs are not 
isomorphic, these lowest cost maps have little inherent interest and so we do 
not list them.

\subsubsection{Isomorphic graphs}
\label{sec4a2}

\begin{figure}[h!]
\begin{center}
\setlength{\unitlength}{0.05in}
\begin{picture}(90,15)(0,-1)
\thicklines
\put(10,12.5){\circle*{1}}
\put(20,12.5){\circle*{1}}
\put(10,2.5){\circle*{1}}
\put(20,2.5){\circle*{1}}
\put(10,2.5){\line(0,1){10}}
\put(20,2.5){\line(0,1){10}}
\put(10,2.5){\line(1,0){10}}
\put(10,12.5){\line(1,0){10}}
\put(10,12.5){\line(1,-1){10}}
\put(8.0,13){$0$}
\put(8.0,1.0){$3$}
\put(21.2,13){$1$}
\put(21.2,1.0){$2$}
\put(14.5,-1){\textbf{\underline{G}}}

\put(47.5,12.5){\circle*{1}}
\put(57.5,12.5){\circle*{1}}
\put(47.5,2.5){\circle*{1}}
\put(57.5,2.5){\circle*{1}}
\put(45.5,13){$0$}
\put(45.5,1.0){$3$}
\put(58.7,13){$1$}
\put(58.7,1.0){$2$}
\put(47.5,2.5){\line(0,1){10}}
\put(47.5,2.5){\line(1,0){10}}
\put(47.5,12.5){\line(1,0){10}}
\put(47.5,12.5){\line(1,-1){10}}
\put(47.5,2.5){\line(1,1){10}}
\put(52,-1){$\mathbf{\underline{G}^{\prime}}$}
\end{picture}
\end{center}
\caption{\label{fig2}Two isomorphic $4$-vertex graphs $G$ and $G^{\prime}$.}
\end{figure}
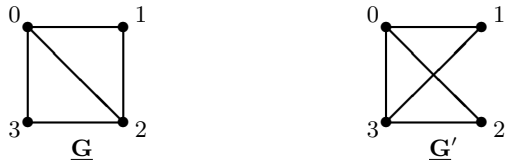

Here we examine the case of two isomorphic graphs $G$ and $G^{\prime}$ which 
are shown in Figure~\ref{fig2}.
Each graph contains $4$ vertices and $5$ edges, and both graphs have
degree sequence $\{ 3,3,2,2\}$. By inspection of Figure~\ref{fig2}, the 
associated adjacency matrices are, respectively,
\begin{equation}
A = \left(
                \begin{array}{cccc}
                   0 & 1 & 1 & 1 \\
                   1 & 0 & 1 & 0 \\
                   1 & 1 & 0 & 1 \\
                   1 & 0 & 1 & 0 \\
               \end{array}
      \right) 
\hspace{0.15in} ; \hspace{0.15in}  
A^{\prime} = \left(
                               \begin{array}{cccc}
                                  0 & 1 & 1 & 1 \\
                                  1 & 0 & 0 & 1 \\
                                  1 & 0 & 0 & 1 \\
                                  1 & 1 & 1 & 0 \\
                               \end{array}
                       \right) .
\label{twoadmat2}
\end{equation}
For these two graphs, the GI AQA finds a vanishing final ground-state energy
$E_{gs} = 0$ and so recognizes $G$ and $G^{\prime}$ as isomorphic. It also finds
that the final ground-state subspace has a degeneracy of $4$. The four CBS that
span this subspace give four binary strings $s_{b}$. These four binary strings 
in turn generate four integer strings $s$ which are, respectively, the bottom 
row of four permutations (see Eq.~(\ref{permdef})). Thus, not only does the GI 
AQA recognize $G$ and $G^{\prime}$ as isomorphic, but it also returns the $4$ 
graph isomorphisms $\pi_{1},\ldots ,\pi_{4}$ that transform $G\rightarrow 
G^{\prime}$ while preserving adjacency:
\begin{eqnarray}
\pi_{1} = \left( \begin{array}{cccc}
                                  0 & 1 & 2 & 3 \\
                                  0 & 2 & 3 & 1 \\
                               \end{array}
                     \right) 
 & ; & \hspace{0.05in}
\pi_{2} = \left( \begin{array}{cccc}
                                  0 & 1 & 2 & 3 \\
                                  3 & 2 & 0 & 1 \\
                               \end{array}
                     \right)  ; \nonumber\\
\pi_{3} = \left( \begin{array}{cccc}
                                  0 & 1 & 2 & 3 \\
                                  3 & 1 & 0 & 2 \\
                               \end{array}
                     \right) 
 & ; & \hspace{0.05in}
\pi_{4} = \left( \begin{array}{cccc}
                                  0 & 1 & 2 & 3 \\
                                  0 & 1 & 3 & 2 \\
                               \end{array}
                     \right)  .
\label{perms1}
\end{eqnarray}
We will next explicitly show that $\pi_{1}$ is a graph isomorphism; the reader
can easily check that the remaining $3$ permutations are also graph 
isomorphisms.

\begin{figure}[h!]
\begin{center}
\setlength{\unitlength}{0.05in}
\begin{picture}(90,15)(0,-1)
\thicklines
\put(10,12.5){\circle*{1}}
\put(20,12.5){\circle*{1}}
\put(10,2.5){\circle*{1}}
\put(20,2.5){\circle*{1}}
\put(10,2.5){\line(0,1){10}}
\put(20,2.5){\line(0,1){10}}
\put(10,2.5){\line(1,0){10}}
\put(10,12.5){\line(1,0){10}}
\put(10,12.5){\line(1,-1){10}}
\put(8,13){$0$}
\put(8,1.0){$3$}
\put(21.2,13){$1$}
\put(21.2,1.0){$2$}
\put(14.5,-1){\textbf{\underline{G}}}

\put(32,7.5){$\longrightarrow$}

\put(47.5,12.5){\circle*{1}}
\put(57.5,12.5){\circle*{1}}
\put(47.5,2.5){\circle*{1}}
\put(57.5,2.5){\circle*{1}}
\put(45.5,13){$0$}
\put(45.5,1.0){$1$}
\put(58.7,13){$2$}
\put(58.7,1.0){$3$}
\put(47.5,2.5){\line(0,1){10}}
\put(47.5,2.5){\line(1,0){10}}
\put(47.5,12.5){\line(1,0){10}}
\put(47.5,12.5){\line(1,-1){10}}
\put(57.5,2.5){\line(0,1){10}}
\put(49,-1){$\mathbf{\underline{\pi_{1}(G)}}$}
\end{picture}
\end{center}
\caption{\label{fig3}Transformation of $G$ produced by the permutation 
$\pi_{1}$.}
\end{figure}
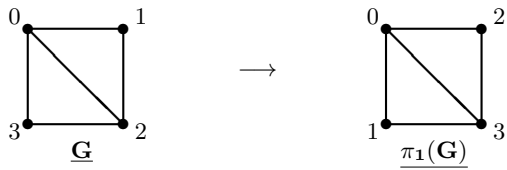

The permutation matrix $\sigma_{1}$ associated with $\pi_{1}$ is
\begin{equation}
\sigma_{1} = \left( \begin{array}{cccc}
                                    1 & 0 & 0 & 0 \\
                                    0 & 0 & 0 & 1 \\
                                    0 & 1 & 0 & 0 \\
                                    0 & 0 & 1 & 0 \\
                               \end{array}
                      \right) .
\label{sigma1mat}
\end{equation}
Under $\pi_{1}$, $G$ is transformed to $\pi_{1}(G)$ which is shown in 
Figure~\ref{fig3}.
It is clear from  Figure~\ref{fig3} that vertices $x$ and $y$ are adjacent in
$G$ if and only if $\pi_{1,x}$ and $\pi_{1,y}$ are adjacent in $\pi_{1}(G)$. 
The adjacency matrix for $\pi_{1}(G)$ is $\sigma_{1}A\sigma_{1}^{T}$ which is
easily shown to be
\begin{equation}
\sigma_{1}A\sigma_{1}^{T} = \left( \begin{array}{cccc}
                                                               0 & 1 & 1 & 1 \\
                                                               1 & 0 & 0 & 1 \\
                                                               1 & 0 & 0 & 1 \\
                                                               1 & 1 & 1 & 0 \\
                                                           \end{array}
                                                  \right) .
\label{newadmat}
\end{equation}
This agrees with the adjacency shown in $\pi_{1}(G)$ in Figure~\ref{fig3}.
Comparison with Eq.~(\ref{twoadmat2}) shows that $\sigma_{1}A\sigma_{1}^{T}
= A^{\prime}$. Thus the permutation $\pi_{1}$ does map $G$ into $G^{\prime}$ 
and preserve adjacency and so establishes that $G$ and $G^{\prime}$ are 
isomorphic.

Finally, note that $G$ and $G^{\prime}$ are connected by exactly $4$ 
graph isomorphisms. Examination of Figure~\ref{fig2} shows that the degree-$3$
vertices in $G$ are vertices $0$ and $2$, while in $G^{\prime}$ they are 
vertices $0$ and $3$. Since the degree of a vertex is preserved by a graph
isomorphism \cite{Chart&Zhang}, vertex $0$ in $G$ must be mapped to vertex 
$0$ or $3$ in $G^{\prime}$. Then vertex $2$ (in $G$) must be mapped to vertex 
$3$ or $0$ (in $G^{\prime}$), respectively. This then forces vertex $1$ to map 
to vertex $1$ or $2$, and vertex $3$ to map to vertex $2$ or $1$, respectively. 
Thus only $4$ graph isomorphism are possible and these are exactly the $4$ 
graph isomorphisms found by the GI AQA which appear in Eq.~(\ref{perms1}).

\subsection{Non-isomorphic graphs}
\label{sec4b}

In this subsection we present GI instances involving pairs of non-isomorphic
graphs. In Section~\ref{sec4b1} we examine two instances of  isospectral graphs;
and in Section~\ref{sec4b2} we look at three instances of strongly regular 
graphs. We shall see that the GI AQA correctly distinguishes all graph pairs as 
non-isomorphic.

\subsubsection{Iso-spectral graphs}
\label{sec4b1}

The spectrum of a graph is the set containing all the eigenvalues of its
adjacency matrix. Two graphs are iso-spectral if they have identical spectra. 
Non-isomorphic iso-spectral graphs are believed to be difficult to distinguish 
\cite{Shiau,Gamb}. Here we test the GI AQA on pairs of non-isomorphic 
iso-spectral graphs. The non-isomorphic pairs of iso-spectral 
graphs examined here appear in Ref.~\cite{isospec}. \\

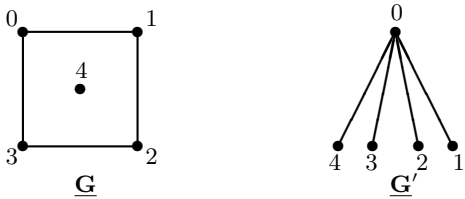
\begin{figure}[h!]
\begin{center}
\setlength{\unitlength}{0.06in}
\begin{picture}(90,15)(0,-2)
\thicklines
\put(10,12.5){\circle*{1}}
\put(10,2.5){\circle*{1}}
\put(20,12.5){\circle*{1}}
\put(20,2.5){\circle*{1}}
\put(15,7.5){\circle*{1}}
\put(10,2.5){\line(0,1){10}}
\put(20,2.5){\line(0,1){10}}
\put(10,2.5){\line(1,0){10}}
\put(10,12.5){\line(1,0){10}}
\put(8.5,13){$0$}
\put(8.5,1){$3$}
\put(20.7,13){$1$}
\put(20.7,1){$2$}
\put(14.6,8.5){$4$}
\put(14.5,-1.5){\textbf{\underline{G}}}

\put(42.5,12.5){\circle*{1}}
\put(37.5,2.5){\circle*{1}}
\put(47.5,2.5){\circle*{1}}
\put(40.5,2.5){\circle*{1}}
\put(44.5,2.5){\circle*{1}}
\put(37.5,2.5){\line(1,2){5}}
\put(47.5,2.5){\line(-1,2){5}}
\put(40.5,2.5){\line(1,5){2}}
\put(44.5,2.5){\line(-1,5){2}}
\put(42.1,13.5){$0$}
\put(47.5,0.4){$1$}
\put(44.3,0.4){$2$}
\put(39.9,0.4){$3$}
\put(36.7,0.4){$4$}
\put(42,-1.5){$\mathbf{\underline{G}^{\prime}}$}

\end{picture}
\end{center}
\caption{\label{fig4}Two non-isomorphic $5$-vertex iso-spectral graphs
$G$ and $G^{\prime}$.}
\end{figure}

\underline{$N = 5:$} It is known that no pair of graphs with less than $5$
vertices is iso-spectral \cite{isospec}. Figure~\ref{fig4} 
shows a pair of graphs $G$ and $G^{\prime}$ with $5$ vertices which are 
isospectral and yet are non-isomorphic. Although both have $5$ vertices and 
$4$ edges, the degree sequence of $G$ is $\{ 2,2,2,2,0\}$, while that of 
$G^{\prime}$ is $\{ 4,1,1,1,1\}$. Since they have different degree sequences, 
it follows that $G$ and $G^{\prime}$ are non-isomorphic. The adjacency matrices 
$A$ and $A^{\prime}$ for $G$ and $G^{\prime}$, respectively, are:
\begin{equation}
A = \left(  \begin{array}{ccccc}
                    0 & 1 & 0 & 1 & 0 \\
                    1 & 0 & 1 & 0 & 0 \\
                    0 & 1 & 0 & 1 & 0 \\
                    1 & 0 & 1 & 0 & 0 \\
                    0 & 0 & 0 & 0 & 0 \\
                 \end{array}
       \right)
\hspace{0.05in} ; \hspace{0.05in}
A^{\prime} =  \left(  \begin{array}{ccccc}
                                     0 & 1 & 1 & 1 & 1 \\
                                     1 & 0 & 0 & 0 & 0 \\
                                     1 & 0 & 0 & 0 & 0 \\
                                     1 & 0 & 0 & 0 & 0 \\
                                     1 & 0 & 0 & 0 & 0 \\
                                  \end{array}
                       \right) .
\label{admatiso1}
\end{equation}
It is straight-forward to show that $A$ and $A^{\prime}$ have the same
characteristic polynomial $P(\lambda ) = \lambda^{5} - 4\lambda^{3}$ and
so $G$ and $G^{\prime}$ are iso-spectral. Numerical simulation of the dynamics 
of the GI AQA finds a non-zero final ground-state energy $E_{gs}= 5$, and so 
the GI AQA correctly distinguishes $G$ and $G^{\prime}$ as non-isomorphic 
graphs.\\

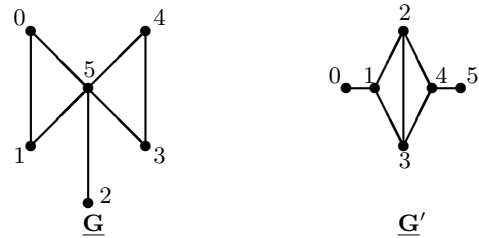
\begin{figure}[h!]
\begin{center}
\setlength{\unitlength}{0.06in}
\begin{picture}(90,20)(0,-5)
\thicklines
\put(10,12.5){\circle*{1}}
\put(10,2.5){\circle*{1}}
\put(20,12.5){\circle*{1}}
\put(20,2.5){\circle*{1}}
\put(15,7.5){\circle*{1}}
\put(15,-2.5){\circle*{1}}
\put(10,2.5){\line(0,1){10}}
\put(20,2.5){\line(0,1){10}}
\put(10,2.5){\line(1,1){5}}
\put(10,12.5){\line(1,-1){5}}
\put(15,7.5){\line(1,1){5}}
\put(15,7.5){\line(1,-1){5}}
\put(15,-2.5){\line(0,1){10}}
\put(8.5,13){$0$}
\put(8.5,1){$1$}
\put(16,-2.5){$2$}
\put(20.7,13){$4$}
\put(20.7,1){$3$}
\put(14.6,8.5){$5$}
\put(14.5,-5){\textbf{\underline{G}}}

\put(37.5,7.5){\circle*{1}}
\put(40,7.5){\circle*{1}}
\put(47.5,7.5){\circle*{1}}
\put(45,7.5){\circle*{1}}
\put(42.5,12.5){\circle*{1}}
\put(42.5,2.5){\circle*{1}}
\put(37.5,7.5){\line(1,0){2.5}}
\put(45,7.5){\line(1,0){2.5}}
\put(42.5,2.5){\line(0,1){10}}
\put(40,7.5){\line(1,-2){2.5}}
\put(42.5,2.5){\line(1,2){2.5}}
\put(40,7.5){\line(1,2){2.5}}
\put(42.5,12.5){\line(1,-2){2.5}}

\put(36,8){$0$}
\put(39,8){$1$}
\put(42.1,13.5){$2$}
\put(42.1,0.5){$3$}
\put(45.3,8){$4$}
\put(48,8){$5$}
\put(42,-5){$\mathbf{\underline{G}^{\prime}}$}

\end{picture}
\end{center}
\caption{\label{fig5}Two non-isomorphic $6$-vertex iso-spectral graphs
$G$ and $G^{\prime}$.}
\end{figure}

\underline{$N = 6:$} The pair of graphs $G$ and $G^{\prime}$ in 
Figure~\ref{fig5}
are iso-spectral and non-isomorphic. To see this, note that although both graphs
have $6$ vertices and $7$ edges, the degree sequence of $G$ is 
$\{ 5,2,2,2,2,1\}$, while that of $G^{\prime}$ is $\{ 3,3,3,3,1,1\}$. Since
their degree sequences are different, $G$ and $G^{\prime}$ are non-isomorphic.
The adjacency matrices $A$ and $A^{\prime}$ of $G$ and $G^{\prime}$, 
respectively, are:
\begin{equation}
A = \left(  \begin{array}{cccccc}
                    0 & 1 & 0 & 0 & 0 & 1 \\
                    1 & 0 & 0 & 0 & 0 & 1 \\
                    0 & 0 & 0 & 0 & 0 & 1 \\
                    0 & 0 & 0 & 0 & 1 & 1 \\
                    0 & 0 & 0 & 1 & 0 & 1 \\
                    1 & 1 & 1 & 1 & 1 & 0 \\
                 \end{array}
       \right)
\hspace{0.05in} ; \hspace{0.05in}
A^{\prime} =  \left(  \begin{array}{cccccc}
                                     0 & 1 & 0 & 0 & 0 & 0 \\
                                     1 & 0 & 1 & 1 & 0 & 0 \\
                                     0 & 1 & 0 & 1 & 1 & 0 \\
                                     0 & 1 & 1 & 0 & 1 & 0 \\
                                     0 & 0 & 1 & 1 & 0 & 1 \\
                                     0 & 0 & 0 & 0 & 1 & 0 \\
                                  \end{array}
                       \right) .
\label{admatiso2}
\end{equation}
Both $A$ and $A^{\prime}$ have the same characteristic polynomial $P(\lambda )
= \lambda^{6} - 7\lambda^{4} - 4\lambda^{3} + 7\lambda^{2} + 4\lambda -1$
which establishes that $G$ and $G^{\prime}$ are iso-spectral. Numerical 
simulation of the dynamics of the GI AQA finds a non-zero final ground-state 
energy $E_{gs} = 7$, and so the GI AQA correctly distinguishes $G$ and 
$G^{\prime}$ as non-isomorphic graphs.

\subsubsection{Strongly regular graphs}
\label{sec4b2}

As we saw in Section~\ref{sec2a}, the degree $d(x)$ of a vertex $x$ is equal
to the number of vertices that are adjacent to $x$. A graph $G$ is said to be 
$k$-\textit{regular\/} if, for all vertices $x$, the degree $d(x) = k$. A graph 
is said to be \textit{regular\/} if it is $k$-regular for some value of $k$. 
Finally, a \textit{strongly regular\/} graph is a graph with $\nu$ vertices 
that is $k$-regular, and for which: (i)~any two adjacent vertices have 
$\lambda$ common neighbors; and (ii)~any two non-adjacent vertices have $\mu$ 
common neighbors. The set of all strongly regular graphs is divided up into 
families, and each family is composed of strongly regular graphs having the
same parameter values $(\nu , k, \lambda ,\mu )$. In this subsection we apply 
the GI AQA to pairs of non-isomorphic strongly regular graphs. An excellent 
test for this algorithm would be two non-isomorphic strongly regular graphs 
belonging to the \textit{same\/} family as such graphs would then have the 
same order and degree sequence. Unfortunately, to find a family containing 
\textit{at least\/} two non-isomorphic strongly regular graphs requires going 
to a family containing $16$-vertex graphs. For example, the family $(16,9,4,6)$ 
contains two non-isomorphic strongly regular graphs. Since the GI AQA uses 
$\nu\lceil\log_{2}\nu\rceil$ qubits, simulation of its quantum dynamics on 
$16$-vertex graphs requires $64$ qubits. This is hopelessly beyond the 
$20$--$22$ qubit limit for such simulations discussed 
in the introduction to Section~\ref{sec4}. As noted there, this hard limit 
restricts our simulations to graphs with no more than $7$ vertices. Now the 
number of connected strongly regular graphs with $\nu = 4,5,6,7$ is $3,2,5,1$, 
respectively. For each of  the values $\nu = 4,5,6$, the strongly regular 
graphs are non-isomorphic as desired, however, each graph belongs to a
\textit{different\/} family. In light of the above remarks, the simulations 
reported in this subsection are restricted to pairs of connected non-isomorphic
strongly regular graphs with $\nu = 4,5,6$.  Ideally we would have simulated 
the GI AQA on the above pair of $16$-vertex strongly regular graphs, however 
the realities of simulating quantum systems on a classical computer made this 
test well beyond reach.\\

\begin{figure}[h!]
\begin{center}
\setlength{\unitlength}{0.05in}
\begin{picture}(90,16)(0,-2)
\thicklines
\put(15,12.5){\circle*{1}}
\put(10,7.5){\circle*{1}}
\put(15,2.5){\circle*{1}}
\put(20,7.5){\circle*{1}}
\put(10,7.5){\line(1,1){5}}
\put(20,7.5){\line(-1,1){5}}
\put(10,7.5){\line(1,-1){5}}
\put(20,7.5){\line(-1,-1){5}}

\put(14.7,13.5){$0$}
\put(8,8){$3$}
\put(20.7,8){$1$}
\put(15.7,0.7){$2$}
\put(13.5,-3){\textbf{\underline{G}}}

\put(52.5,12.5){\circle*{1}}
\put(52.5,2.5){\circle*{1}}
\put(47.5,7.5){\circle*{1}}
\put(57.5,7.5){\circle*{1}}

\put(52.2,13.5){$0$}
\put(45.5,8){$3$}
\put(58.2,8){$1$}
\put(53.2,0.7){$2$}

\put(47.5,7.5){\line(1,1){5}}
\put(47.5,7.5){\line(1,-1){5}}
\put(52.5,12.5){\line(1,-1){5}}
\put(52.5,2.5){\line(1,1){5}}
\put(47.5,7.5){\line(1,0){10}}
\put(52.5,2.5){\line(0,1){10}}
\put(51,-3){$\mathbf{\underline{G}^{\prime}}$}
\end{picture}
\end{center}
\caption{\label{fig6}Two $4$-vertex non-isomorphic strongly regular graphs 
$G$ and $G^{\prime}$.}
\end{figure}
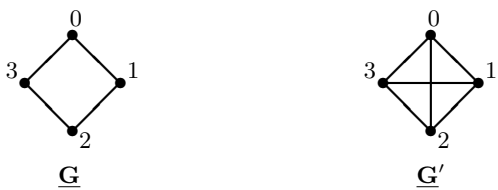

\underline{$N=4:$} In Figure~\ref{fig6} 
we show two strongly regular $4$-vertex graphs $G$ and $G^{\prime}$. The 
parameters for $G$ are $\nu = 4$, $k=2$, $\lambda = 0$, and $\mu = 2$; while 
for $G^{\prime}$ they are $\nu = 4$, $k=3$, $\lambda = 2$, and $\mu = 0$. It 
is clear that $G$ and $G^{\prime}$ are non-isomorphic since they contain an
unequal number of edges and different degree sequences. The adjacency matrices
$A$ and $A^{\prime}$ for $G$ and $G^{\prime}$ are, respectively,
\begin{equation}
A = \left(  \begin{array}{cccc}
                     0 & 1 & 0 & 1 \\
                     1 & 0 & 1 & 0 \\
                     0 & 1 & 0 & 1 \\
                     1 & 0 & 1 & 0 \\
                \end{array}
       \right)
\hspace{0.05in} ; \hspace{0.05in}
A^{\prime} = \left(  \begin{array}{cccc}
                                     0 & 1 & 1 & 1 \\
                                     1 & 0 & 1 & 1 \\
                                     1 & 1 & 0 & 1 \\
                                     1 & 1 & 1 & 0 \\
                                \end{array}
                      \right) .
\label{strreggrphs1}
\end{equation}
Numerical simulation of the GI AQA dynamics finds a non-zero final 
ground-state energy $E_{gs} = 4$ and so the GI AQA correctly distinguishes
$G$ and $G^{\prime}$ as non-isomorphic. \\

\begin{figure}[h!]
\begin{center}
\setlength{\unitlength}{0.05in}
\begin{picture}(90,18)(0,-3)
\thicklines
\put(10,7.5){\circle*{1}}
\put(10,2.5){\circle*{1}}
\put(20,7.5){\circle*{1}}
\put(20,2.5){\circle*{1}}
\put(15,12.5){\circle*{1}}

\put(10,2.5){\line(0,1){5}}
\put(20,2.5){\line(0,1){5}}
\put(10,2.5){\line(1,0){10}}
\put(10,7.5){\line(1,1){5}}
\put(20,7.5){\line(-1,1){5}}

\put(14.6,13.5){$0$}
\put(8,1){$3$}
\put(21.2,8){$1$}
\put(21.2,1){$2$}
\put(8,8){$4$}

\put(14.5,-3){\textbf{\underline{G}}}

\put(57.5,2.5){\circle*{1}}
\put(47.5,7.5){\circle*{1}}
\put(57.5,7.5){\circle*{1}}
\put(47.5,2.5){\circle*{1}}
\put(52.5,12.5){\circle*{1}}

\put(47.5,7.5){\line(1,1){5}}
\put(57.5,7.5){\line(-1,1){5}}
\put(47.5,2.5){\line(1,2){5}}
\put(57.5,2.5){\line(-1,2){5}}
\put(47.5,2.5){\line(0,1){5}}
\put(57.5,2.5){\line(0,1){5}}
\put(47.5,2.5){\line(1,0){10}}
\put(47.5,7.5){\line(1,0){10}}
\put(47.5,2.5){\line(2,1){10}}
\put(47.5,7.5){\line(2,-1){10}}

\put(52.1,13.5){$0$}
\put(58.7,8){$1$}
\put(58.7,1){$2$}
\put(45.5,1){$3$}
\put(45.5,8){$4$}

\put(52,-3){$\mathbf{\underline{G}^{\prime}}$}

\end{picture}
\end{center}
\caption{\label{fig7}Two $5$-vertex non-isomorphic strongly regular graphs
$G$ and $G^{\prime}$.}
\end{figure}
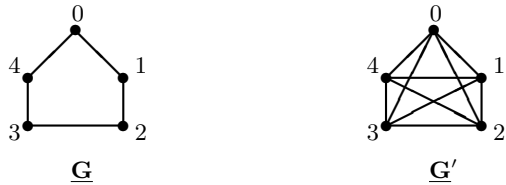

\underline{$N=5:$} In Figure~\ref{fig7}
we show two strongly regular $5$-vertex graphs $G$ and $G^{\prime}$. The 
parameters for $G$ are $\nu = 5$, $k=2$, $\lambda = 0$, and $\mu = 1$; while
for $G^{\prime}$ they are $\nu = 5$, $k=4$, $\lambda = 3$, and $\mu = 0$. It 
is clear that $G$ and $G^{\prime}$ are non-isomorphic since they contain an
unequal number of edges and different degree sequences. The adjacency matrices
$A$ and $A^{\prime}$ for $G$ and $G^{\prime}$ are, respectively,
\begin{equation}
A = \left(  \begin{array}{ccccc}
                     0 & 1 & 0 & 0 & 1 \\
                     1 & 0 & 1 & 0 & 0 \\
                     0 & 1 & 0 & 1 & 0 \\
                     0 & 0 & 1 & 0 & 1 \\
                     1 & 0 & 0 & 1 & 0 \\
                \end{array}
       \right)
\hspace{0.05in} ; \hspace{0.05in}
A^{\prime} = \left(  \begin{array}{ccccc}
                                     0 & 1 & 1 & 1 & 1 \\
                                     1 & 0 & 1 & 1 & 1 \\
                                     1 & 1 & 0 & 1 & 1 \\
                                     1 & 1 & 1 & 0 & 1 \\
                                     1 & 1 & 1 & 1 & 0 \\
                                \end{array}
                      \right) .
\label{strreggrphs2}
\end{equation}
Numerical simulation of the GI AQA dynamics finds a non-zero final 
ground-state energy $E_{gs} = 10$ and so the GI AQA correctly distinguishes
$G$ and $G^{\prime}$ as non-isomorphic. \\

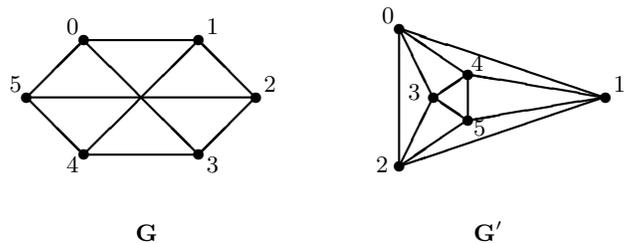
\begin{figure}[h!]
\begin{center}
\setlength{\unitlength}{0.06in}
\begin{picture}(90,20)(0,-5)
\thicklines
\put(10,12.5){\circle*{1}}
\put(10,2.5){\circle*{1}}
\put(20,12.5){\circle*{1}}
\put(20,2.5){\circle*{1}}
\put(5,7.5){\circle*{1}}
\put(25,7.5){\circle*{1}}

\put(10,2.5){\line(1,1){10}}
\put(10,12.5){\line(1,-1){10}}
\put(10,2.5){\line(1,0){10}}
\put(10,12.5){\line(1,0){10}}
\put(5,7.5){\line(1,1){5}}
\put(5,7.5){\line(1,-1){5}}
\put(25,7.5){\line(-1,1){5}}
\put(25,7.5){\line(-1,-1){5}}
\put(5,7.5){\line(1,0){20}}

\put(8.5,13){$0$}
\put(8.5,1){$4$}
\put(20.7,13){$1$}
\put(20.7,1){$3$}
\put(25.7,8){$2$}
\put(3.5,8){$5$}
\put(14.5,-5){\textbf{\underline{G}}}

\put(40.5,7.5){\circle*{1}}
\put(38.3,7.3){$3$}

\put(43.5,9.5){\circle*{1}}
\put(43.8,9.8){$4$}

\put(55.5,7.5){\circle*{1}}
\put(56.2,8){$1$}

\put(43.5,5.5){\circle*{1}}
\put(44,4.2){$5$}

\put(37.5,13.5){\circle*{1}}
\put(36,14){$0$}

\put(37.5,1.5){\circle*{1}}
\put(35.5,1){$2$}

\put(37.5,1.5){\line(1,2){3}}
\put(37.5,13.5){\line(1,-2){3}}
\put(37.5,1.5){\line(0,1){12}}
\put(40.5,7.5){\line(3,2){3}}
\put(40.5,7.5){\line(3,-2){3}}
\put(43.5,5.5){\line(0,1){4}}
\put(43.5,9.5){\line(-3,2){6}}
\put(43.5,5.5){\line(-3,-2){6}}
\put(55.5,7.5){\line(-6,1){12}}
\put(55.5,7.5){\line(-6,-1){12}}
\put(55.5,7.5){\line(-3,1){18}}
\put(55.5,7.5){\line(-3,-1){18}}

\put(44,-5){$\mathbf{\underline{G}^{\prime}}$}

\end{picture}
\end{center}
\caption{\label{fig8}Two $6$-vertex non-isomorphic strongly regular graphs
$G$ and $G^{\prime}$.}
\end{figure}

\underline{$N=6:$} In Figure~\ref{fig8}
we show two strongly regular $6$-vertex graphs $G$ and $G^{\prime}$. The 
parameters for $G$ are $\nu = 6$, $k=3$, $\lambda = 0$, and $\mu = 3$; while
for $G^{\prime}$ they are $\nu = 6$, $k=4$, $\lambda = 2$, and $\mu = 4$. It 
is clear that $G$ and $G^{\prime}$ are non-isomorphic since they contain an
unequal number of edges and different degree sequences. The adjacency matrices
$A$ and $A^{\prime}$ for $G$ and $G^{\prime}$ are, respectively,
\begin{equation}
A = \left(  \begin{array}{cccccc}
                     0 & 1 & 0 & 1 & 0 & 1 \\
                     1 & 0 & 1 & 0 & 1 & 0 \\
                     0 & 1 & 0 & 1 & 0 & 1 \\
                     1 & 0 & 1 & 0 & 1 & 0 \\
                     0 & 1 & 0 & 1 & 0 & 1 \\
                     1 & 0 & 1 & 0 & 1 & 0 \\
                \end{array}
       \right)
\hspace{0.0in} ; \hspace{0.05in}
A^{\prime} = \left(  \begin{array}{cccccc}
                                     0 & 1 & 1 & 1 & 1 & 0 \\
                                     1 & 0 & 1 & 0 & 1 & 1 \\
                                     1 & 1 & 0 & 1 & 0 & 1 \\
                                     1 & 0 & 1 & 0 & 1 & 1 \\
                                     1 & 1 & 0 & 1 & 0 & 1 \\
                                     0 & 1 & 1 & 1 & 1 & 0 \\
                                \end{array}
                      \right) .
\label{strreggrphs3}
\end{equation}
Numerical simulation of the GI AQA dynamics finds a non-zero final 
ground-state energy $E_{gs} = 10$ and so the GI AQA correctly distinguishes
$G$ and $G^{\prime}$ as non-isomorphic. 

\subsection{Graph automorphisms}
\label{sec4c}

As noted in the Section~\ref{sec4} introduction, we can find the automorphism 
group $Aut(G)$ of a graph $G$ using the GI AQA by considering a GI instance 
with $G^{\prime} = G$. To the best of our knowledge, such an application of 
a GI algorithm is new. Here the self-isomorphisms are permutations of the 
vertices of $G$ that map $G\rightarrow G$ while preserving adjacency. Since 
$G$ is always isomorphic to itself, the final ground-state energy will vanish:
$E_{gs}=0$. The set of CBS $|s_{b}\rangle$ that span the final ground-state 
subspace give rise to a set of binary strings $s_{b}$ that determine the 
integer strings $s=s_{0}\cdots s_{N-1}$ (see Section~\ref{sec2b}) that then 
determine the permutations $\pi (s)$,
\begin{equation}
\pi (s) = \left(  \begin{array}{ccccc}
                            0 & \cdots & i & \cdots & N-1 \\
                           s_{0} & \cdots & s_{i} & \cdots & s_{N-1} \\
                        \end{array}
              \right) ,
\label{permdefsec4c}
\end{equation}
which are all the elements of $Aut(G)$. By construction, the order of $Aut(G)$ 
is equal to the degeneracy of the final ground-state subspace. In this 
subsection we apply the GI AQA to the: (i)~cycle graphs $C_{4}, \ldots , 
C_{7}$; (ii)~grid graph $G_{2,3}$; and (iii)~wheel graph $W_{7}$, and show
that it correctly determines the automorphism group for all of these graphs.

\subsubsection{Cycle graphs}
\label{sec4c1}

A \textit{walk\/} $W$ in a graph is an alternating sequence of vertices and
edges $x_{0}$, $e_{1}$, $x_{1}$, $e_{2}$, $\cdots$, $e_{l}$, $x_{l}$, where 
the edge $e_{i}$ connects $x_{i-1}$ and $x_{i}$ for $0<i\leq l$. A walk $W$ is 
denoted by the sequence of vertices it traverses $W = x_{0}x_{1}\cdots x_{l}$.
Finally, a walk $W= x_{0}x_{1}\cdots x_{l}$ is a \textit{cycle\/} if $l\geq 3$;
$x_{0}=x_{l}$; and the vertices $x_{i}$ with $0 < i < l$ are distinct from 
each other and $x_{0}$. A cycle with $n$ vertices is denoted $C_{n}$. 

The automorphism group of the cycle graph $C_{n}$ is the dihedral group $D_{n}$
\cite{Chart&Zhang}. The order of $D_{n}$ is $2n$, and it is generated by the two 
elements $\alpha$ and $\beta$ that satisfy the following relations,
\begin{equation}
\alpha^{n} = e ; \hspace{0.5in} \beta^{2} = e ; \hspace{0.5in} 
                \alpha\beta = \beta\alpha^{n-1} ,
\label{dihedralrels}
\end{equation}
where $e$ is the identity element. Because $\alpha$ and $\beta$ are generators
of $D_{n}$, each element $g$ of $D_{n}$ can be written as a product of 
appropriate powers of $\alpha$ and $\beta$:
\begin{equation}
g = \alpha^{i}\beta^{j}  \hspace{0.5in} ( 0\leq i\leq n-1 \hspace{0.1in} ;
               \hspace{0.1in} 0 \leq j\leq 1).
\label{prodgen}
\end{equation}

We now use the GI AQA to find the automorphism group of the cycle graphs 
$C_{n}$ for $4\leq n\leq 7$. We will see that the GI AQA correctly determines
$Aut(C_{n}) = D_{n}$ for these graphs. We will work out $C_{4}$ in detail, 
and then give more abbreviated presentations for the remaining cycle
graphs as their analysis is identical. \\

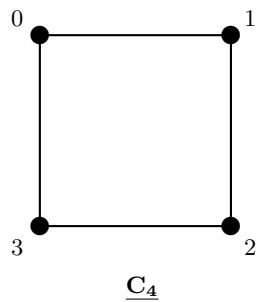
\begin{figure}[h!]
\begin{center}
\setlength{\unitlength}{0.1in}
\begin{picture}(30,17)(0,-1)
\thicklines
\put(10,12.5){\circle*{1}}
\put(10,2.5){\circle*{1}}
\put(20,12.5){\circle*{1}}
\put(20,2.5){\circle*{1}}

\put(10,2.5){\line(0,1){10}}
\put(20,2.5){\line(0,1){10}}
\put(10,2.5){\line(1,0){10}}
\put(10,12.5){\line(1,0){10}}
\put(8.5,13){$0$}
\put(8.5,1){$3$}
\put(20.7,13){$1$}
\put(20.7,1){$2$}

\put(14.5,-1){$\mathbf{\underline{C_{4}}}$}
\end{picture}
\end{center}
\caption{\label{fig9}Cycle graph $C_{4}$.}
\end{figure}

\underline{$N = 4$}: The cycle graph $C_{4}$ appears in Figure~\ref{fig9}.
 It contains $4$ vertices and $4$ edges; has degree sequence $\{ 2,2,2,2\}$; 
and adjacency matrix
\begin{equation}
A = \left(  \begin{array}{cccc}
                     0 & 1 & 0 & 1 \\
                     1 & 0 & 1 & 0 \\
                     0 & 1 & 0 & 1 \\
                     1 & 0 & 1 & 0 \\
                \end{array}
      \right) .
\label{C4admat}
\end{equation}

Numerical simulation of the GI AQA applied to the GI instance with $G = C_{4}$
and $G^{\prime} = G$ found a vanishing final ground-state energy $E_{gs} 
= 0$. The GI AQA thus correctly identifies $C_{4}$ as being isomorphic 
to itself. The simulation also found that the final ground-state subspace is
$8$-fold degenerate. Table~\ref{table1}
\begin{table}[h!]
\caption{\label{table1}Automorphism group $Aut(C_{4})$ of the cycle graph 
$C_{4}$ as found by the GI AQA. The first row lists the integer strings 
$s=s_{0}s_{1}s_{2}s_{3}$
determined by the labels of the eight CBS $|s_{b}\rangle$ that span the final 
ground-state subspace (see text). Each string $s$ determines a graph 
automorphism $\pi (s)$ via Eq.~(\ref{permdefsec4c}). The second row associates 
each integer string $s$ in the first row with a graph automorphism $\pi (s)$. 
It identifies the two strings that give rise to the graph automorphisms 
$\alpha$ and $\beta$ that generate $Aut(C_{4})$, and writes each graph 
automorphism $\pi (s)$ as a product of an appropriate power of $\alpha$ and 
$\beta$. Note that $e$ is the identity automorphism, and the product notation 
assumes the rightmost factor acts first.\\}
\begin{ruledtabular}
\begin{tabular}{c|cccccccc}
$s = s_{0}s_{1}s_{2}s_{3}$ & $3012$  &  $2301$ & $1230$ &  $0123$ &  $0321$
  &  $3210$ & $2103$ & $1032$ \\ \hline
$\pi (s)$ & $\alpha$ &  $\alpha^{2}$ &  $\alpha^{3}$ &  $\alpha^{4} = e$  &  
   $\beta$  &  $\alpha\beta$  &  $\alpha^{2}\beta$  &  $\alpha^{3}\beta$ \\
\end{tabular}
\end{ruledtabular}
\end{table}
lists the integer strings $s = s_{0}s_{1}s_{2}s_{3}$ that result from the 
eight CBS $|s_{b}\rangle$ that span the final ground-state subspace (see
Sections~\ref{sec2b} and \ref{sec3}). Each integer string $s$ determines the
bottom row of a permutation $\pi (s)$ (see eq.~(\ref{permdefsec4c})).  
We explicitly show that $\pi (s)$, for the integer string $s = 3210$, is an 
automorphism of $C_{4}$. The reader can repeat this analysis to show that the 
seven remaining integer strings in Table~\ref{table1} also give rise to 
automorphisms of $C_{4}$. Note that the CBS that span the final ground-state 
subspace determine \textit{all\/} the graph automorphisms of $C_{4}$. This 
follows from the manner in which the GI AQA is constructed since each 
automorphism of $C_{4}$ must give rise to a CBS with vanishing energy, and 
each CBS in the final ground-state subspace gives rise to an automorphism of 
$C_{4}$. These automorphisms form a group $Aut(C_{4})$ and the GI AQA 
has found that the order of $Aut(C_{4})$ is $8$ which is the same as the order 
of the dihedral group $D_{4}$.

The permutation $\pi_{\ast}\equiv \pi(s = 3210)$ is
\begin{equation}
\pi (3210) = \left( \begin{array}{cccc}
                                 0 & 1 & 2 & 3 \\
                                 3 & 2 & 1 & 0 \\
                             \end{array}
                     \right) ,
\end{equation}
and the associated permutation matrix $\sigma_{\ast} \equiv\sigma (3210)$ is
\begin{equation}
\sigma (3210) = \left(  \begin{array}{cccc}
                                         0 & 0 & 0 & 1 \\
                                         0 & 0 & 1 & 0 \\
                                         0 & 1 & 0 & 0 \\
                                         1 & 0 & 0 & 0 \\
                                    \end{array}
                           \right) .
\label{demoC4sigma}
\end{equation}
Under $\pi_{\ast}$, $C_{4}$ is transformed to 
$\pi_{\ast}(C_{4})$ which is shown in Figure~\ref{fig10}.
\begin{figure}[h!]
\begin{center}
\setlength{\unitlength}{0.05in}
\begin{picture}(90,15)(0,-1)
\thicklines
\put(10,12.5){\circle*{1}}
\put(20,12.5){\circle*{1}}
\put(10,2.5){\circle*{1}}
\put(20,2.5){\circle*{1}}
\put(10,2.5){\line(0,1){10}}
\put(20,2.5){\line(0,1){10}}
\put(10,2.5){\line(1,0){10}}
\put(10,12.5){\line(1,0){10}}
\put(8,13){$0$}
\put(8,1.0){$3$}
\put(20.8,13){$1$}
\put(20.8,1.0){$2$}
\put(14,-1){$\mathbf{\underline{C_{4}}}$}

\put(32,7.5){$\mathbf{\longrightarrow}$}

\put(47.5,12.5){\circle*{1}}
\put(57.5,12.5){\circle*{1}}
\put(47.5,2.5){\circle*{1}}
\put(57.5,2.5){\circle*{1}}
\put(45.5,13){$3$}
\put(45.5,1.0){$0$}
\put(58.4,13){$2$}
\put(58.4,1.0){$1$}
\put(47.5,2.5){\line(0,1){10}}
\put(47.5,2.5){\line(1,0){10}}
\put(47.5,12.5){\line(1,0){10}}
\put(57.5,2.5){\line(0,1){10}}
\put(49,-1){$\mathbf{\underline{\pi_{\ast}(C_{4})}}$}
\end{picture}
\end{center}
\caption{\label{fig10}Transformation of $C_{4}$ produced by the permutation 
$\pi_{\ast}$.}
\end{figure}
It is clear from Figure~\ref{fig10} that $x$ and $y$ are adjacent in $C_{4}$
if and only if $\pi_{\ast ,x}$ and $\pi_{\ast ,y}$ are adjacent in $\pi_{\ast}
(C_{4})$. Thus $\pi_{\ast}$ is a permutation of the vertices of $C_{4}$ that
preserves adjacency and so is a graph automorphism of $C_{4}$. We can also 
show this by demonstrating that $\sigma_{\ast} A\sigma_{\ast}^{T}=A$. 
Using Eqs.~(\ref{C4admat}) and (\ref{demoC4sigma}) it is easy to show that
\begin{equation}
\sigma_{\ast} A\sigma_{\ast}^{T} = 
            \left(  \begin{array}{cccc}
                           0 & 1 & 0 & 1 \\
                           1 & 0 & 1 & 0 \\
                           0 & 1 & 0 & 1 \\
                           1 & 0 & 1 & 0 \\
                      \end{array}
            \right) .
\end{equation}
This agrees with the adjacency of edges in $\pi_{\ast}(C_{4})$ appearing 
in Figure~\ref{fig10}, and comparison with Eq.~(\ref{C4admat}) shows that 
$\sigma_{\ast} A \sigma_{\ast}^{T} = A$, confirming that $\pi_{\ast}$ is an 
automorphism of $C_{4}$. 

\begin{figure}[h!]
\begin{center}
\setlength{\unitlength}{0.05in}
\begin{picture}(90,15)(0,-1)
\thicklines
\put(10,12.5){\circle*{1}}
\put(20,12.5){\circle*{1}}
\put(10,2.5){\circle*{1}}
\put(20,2.5){\circle*{1}}
\put(10,2.5){\line(0,1){10}}
\put(20,2.5){\line(0,1){10}}
\put(10,2.5){\line(1,0){10}}
\put(10,12.5){\line(1,0){10}}
\put(8,13){$0$}
\put(8,1.0){$3$}
\put(20.8,13){$1$}
\put(20.8,1.0){$2$}
\put(14,-1){$\mathbf{\underline{C_{4}}}$}

\put(32,7.5){$\mathbf{\longrightarrow}$}

\put(47.5,12.5){\circle*{1}}
\put(57.5,12.5){\circle*{1}}
\put(47.5,2.5){\circle*{1}}
\put(57.5,2.5){\circle*{1}}
\put(45.5,13){$3$}
\put(45.5,1.0){$2$}
\put(58.4,13){$0$}
\put(58.4,1.0){$1$}
\put(47.5,2.5){\line(0,1){10}}
\put(47.5,2.5){\line(1,0){10}}
\put(47.5,12.5){\line(1,0){10}}
\put(57.5,2.5){\line(0,1){10}}
\put(50,-1){$\mathbf{\underline{\alpha (C_{4})}}$}
\end{picture}
\end{center}
\caption{\label{fig11}Transformation of $C_{4}$ produced by the automorphism 
$\alpha$.}
\end{figure}
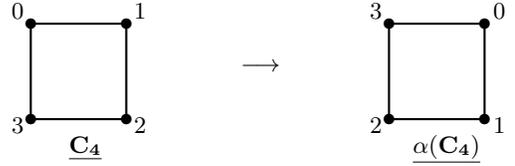

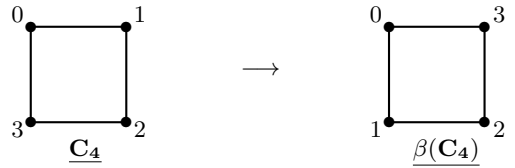
\begin{figure}[ht!]
\begin{center}
\setlength{\unitlength}{0.05in}
\begin{picture}(90,15)(0,-1)
\thicklines
\put(10,12.5){\circle*{1}}
\put(20,12.5){\circle*{1}}
\put(10,2.5){\circle*{1}}
\put(20,2.5){\circle*{1}}
\put(10,2.5){\line(0,1){10}}
\put(20,2.5){\line(0,1){10}}
\put(10,2.5){\line(1,0){10}}
\put(10,12.5){\line(1,0){10}}
\put(8,13){$0$}
\put(8,1.0){$3$}
\put(20.8,13){$1$}
\put(20.8,1.0){$2$}
\put(14,-1){$\mathbf{\underline{C_{4}}}$}

\put(32,7.5){$\mathbf{\longrightarrow}$}

\put(47.5,12.5){\circle*{1}}
\put(57.5,12.5){\circle*{1}}
\put(47.5,2.5){\circle*{1}}
\put(57.5,2.5){\circle*{1}}
\put(45.5,13){$0$}
\put(45.5,1.0){$1$}
\put(58.4,13){$3$}
\put(58.4,1.0){$2$}
\put(47.5,2.5){\line(0,1){10}}
\put(47.5,2.5){\line(1,0){10}}
\put(47.5,12.5){\line(1,0){10}}
\put(57.5,2.5){\line(0,1){10}}
\put(50,-1){$\mathbf{\underline{\beta (C_{4})}}$}
\end{picture}
\end{center}
\caption{\label{fig12}Transformation of $C_{4}$ produced by the automorphism 
$\beta$.}
\end{figure}

We now show that $Aut(C_{4})$ is isomorphic to the dihedral group $D_{4}$
by showing that the automorphisms $\pi (3012) \equiv \alpha$ and $\pi (0321) 
\equiv\beta$ generate $Aut(C_{4})$, and satisfy the generator relations 
(Eq.~(\ref{dihedralrels})) for $D_{4}$. The second row of Table~\ref{table1} 
establishes that $\alpha$ and $\beta$ are the generators of $Aut(C_{4})$ as
it shows that each element of $Aut(C_{4})$ is a product of an appropriate 
power of $\alpha$ and $\beta$, and all possible products of powers of $\alpha$ 
and $\beta$ appear in that row. Now notice that
\begin{equation}
\alpha = \left(   \begin{array}{cccc}
                             0 & 1 & 2 & 3 \\
                             3 & 0 & 1 & 2 \\
                         \end{array}
              \right)
\label{C4alphadef}
\end{equation}
corresponds to a clockwise rotation of $C_{4}$ by $90^{\circ}$ (see 
Figure~\ref{fig11})
Thus four applications of $\alpha$ corresponds to a $360^{\circ}$ rotation
of $C_{4}$ which leaves it invariant. Thus $\alpha^{4} = e$. This can also be 
checked by composing $\alpha$ with itself four times using 
Eq.~(\ref{C4alphadef}). This establishes the 
first of the generator relations in Eq.~(\ref{dihedralrels}). Similarly,
\begin{equation}
\beta = \left(   \begin{array}{cccc}
                             0 & 1 & 2 & 3 \\
                             0 & 3 & 2 & 1 \\
                         \end{array}
              \right)
\end{equation}
corresponds to reflection of $C_{4}$ about the diagonal passing through vertices
$0$ and $2$ (see Figure~\ref{fig12}).
Thus two applications of $\beta$ leaves $C_{4}$ invariant, and so $\beta^{2} =
e$. This establishes the second of the generator relations in 
Eq.~(\ref{dihedralrels}). Finally, to show the third generator relation 
$\alpha\beta = \beta\alpha^{3}$, we simply evaluate both sides of this relation 
and compare results. Using Table~\ref{table1} we find that
\begin{eqnarray}
\alpha\beta & = & \left( \begin{array}{cccc}
                                       0 & 1 & 2 & 3 \\
                                       3 & 0 & 1 & 2 \\
                                   \end{array}
                           \right) 
                           \left( \begin{array}{cccc}
                                       0 & 1 & 2 & 3 \\
                                       0 & 3 & 2 & 1 \\
                                   \end{array}
                           \right)  \nonumber \\
  & = &               \left( \begin{array}{cccc}
                                       0 & 1 & 2 & 3 \\
                                       3 & 2 & 1 & 0 \\
                                   \end{array}
                           \right)  \label{testrel1} \\
\beta\alpha^{3} & = &  \left( \begin{array}{cccc}
                                       0 & 1 & 2 & 3 \\
                                       0 & 3 & 2 & 1 \\
                                   \end{array}
                           \right) 
                           \left( \begin{array}{cccc}
                                       0 & 1 & 2 & 3 \\
                                       1 & 2 & 3 & 0 \\
                                   \end{array}
                           \right)  \nonumber \\
        & = &         \left( \begin{array}{cccc}
                                       0 & 1 & 2 & 3 \\
                                       3 & 2 & 1 & 0 \\
                                   \end{array}
                           \right) . \label{testrel2}
\end{eqnarray}
It is clear that $\alpha\beta$ does equal $\beta\alpha^{3}$. Thus we have shown
that $\alpha$ and $\beta$: (i)~generate $Aut(C_{4})$; and (ii)~satisfy the 
generator relations (Eq.~(\ref{dihedralrels})) for the dihedral group $D_{4}$, 
and so generate a group isomorphic to $D_{4}$. In summary, we have shown that 
the GI AQA found all eight graph automorphisms of $C_{4}$, and that the group 
formed from these automorphisms is isomorphic to the dihedral group $D_{4}$ 
which is the correct automorphism group for $C_{4}$. \\

\begin{figure}[h!]
\begin{center}
\setlength{\unitlength}{0.1in}
\begin{picture}(30,18)(0,-2)
\thicklines
\put(10,7.5){\circle*{1}}
\put(10,2.5){\circle*{1}}
\put(20,7.5){\circle*{1}}
\put(20,2.5){\circle*{1}}
\put(15,12.5){\circle*{1}}

\put(10,2.5){\line(0,1){5}}
\put(20,2.5){\line(0,1){5}}
\put(10,2.5){\line(1,0){10}}
\put(10,7.5){\line(1,1){5}}
\put(20,7.5){\line(-1,1){5}}

\put(14.6,13.5){$0$}
\put(8.5,1){$3$}
\put(20.7,8){$1$}
\put(20.7,1){$2$}
\put(8.5,8){$4$}

\put(14.5,-2){$\mathbf{\underline{C_{5}}}$}
\end{picture}
\end{center}
\caption{\label{fig13}Cycle graph $C_{5}$.}
\end{figure}
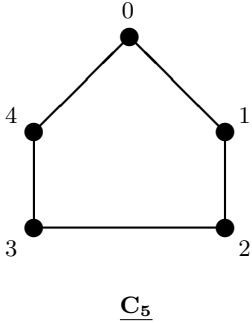

\underline{$N = 5$}: The cycle graph $C_{5}$ appears in Figure~\ref{fig13}.
It has $5$ vertices and $5$ edges; degree sequence $\{ 2,2,2,2,2\}$; and 
adjacency matrix
\begin{equation}
A = \left(  \begin{array}{ccccc}
                     0 & 1 & 0 & 0 & 1 \\
                     1 & 0 & 1 & 0 & 0 \\
                     0 & 1 & 0 & 1 & 0 \\
                     0 & 0 & 1 & 0 & 1 \\
                     1 & 0 & 0 & 1 & 0 \\
                \end{array}
      \right) .
\label{C5admat}
\end{equation}

Numerical simulation of the GI AQA applied to the GI instance with $G = C_{5}$
and $G^{\prime} = G$ found a vanishing final ground-state energy $E_{gs} 
= 0$. The GI AQA thus correctly identifies $C_{5}$ as being isomorphic 
to itself. The simulation also found that the final ground-state subspace is
$10$-fold degenerate. Table~\ref{table2}
\begin{table}[h!]
\caption{\label{table2}Automorphism group $Aut(C_{5})$ of the cycle graph 
$C_{5}$ as found by the GI AQA. The odd rows list the integer strings 
$s=s_{0}\cdots s_{4}$ determined by the labels of the ten CBS $|s_{b}\rangle$ 
that span the final ground-state subspace (see text). Each string $s$ determines
a graph automorphism $\pi (s)$ via Eq.~(\ref{permdefsec4c}). Each even row 
associates each integer string $s$ in the odd row preceding it with a graph 
automorphism $\pi (s)$. It identifies the two strings that give rise to the 
graph automorphisms $\alpha$ and $\beta$ that generate $Aut(C_{5})$, and writes 
each graph automorphism $\pi (s)$ as a product of an appropriate power of 
$\alpha$ and $\beta$. Note that $e$ is the identity automorphism, and the 
product notation assumes the rightmost factor acts first.\\}
\begin{ruledtabular}
\begin{tabular}{c|cccccccccc}
$s = s_{0}\cdots s_{4}$ & $40123$  &  $34012$ & $23401$ &  $12340$ &  $01234$\\
$\pi (s)$ & $\alpha$ &  $\alpha^{2}$ &  $\alpha^{3}$ &  $\alpha^{4}$ & 
    $\alpha^{5} = e$            \\ \hline
$s = s_{0}\cdots s_{4}$ &  $04321$ & $10432$ & $21043$ & $32104$ & $43210$ \\
$\pi (s)$  & $\beta$  &  $\alpha\beta$  &  $\alpha^{2}\beta$  &  
      $\alpha^{3}\beta$ & $\alpha^{4}\beta$ \\
\end{tabular}
\end{ruledtabular}
\end{table}
lists the integer strings $s = s_{0}\cdots s_{4}$ that result from the 
ten CBS $|s_{b}\rangle$ that span the final ground-state subspace (see
Sections~\ref{sec2b} and \ref{sec3}). Each integer string $s$ fixes the
bottom row of a permutation $\pi (s)$ (see eq.~(\ref{permdefsec4c})) which is
a graph automorphism of $C_{5}$. The demonstration of this is identical to the 
demonstration given for $C_{4}$ and so will not be repeated here. Just as for
$Aut(C_{4})$, the graph automorphisms in Table~\ref{table2} are all the elements
of $Aut(C_{5})$ which is seen to have order $10$. Note that this is the same as
the order of  the dihedral group $D_{5}$.

We now show that $Aut(C_{5})$ is isomorphic to the dihedral group $D_{5}$ by
showing that the graph automorphisms $\alpha = \pi (40123)$ and $\beta =
\pi (04321)$ generate $Aut(C_{5})$, and satisfy the generator relations
(Eq.~(\ref{dihedralrels})) for $D_{5}$. The second row of Table~\ref{table2} 
establishes that $\alpha$ and $\beta$ are the generators of $Aut(C_{5})$ as
it shows that each element of $Aut(C_{5})$ is a product of an appropriate 
power of $\alpha$ and $\beta$, and that all possible products of powers of 
$\alpha$ and $\beta$ appear in that row. Following the discussion for $C_{4}$, 
it is a simple matter to show that $\alpha$ corresponds to a $72^{\circ}$ 
clockwise rotation of $C_{5}$. Thus $5$ applications of $\alpha$ rotates $C_{5}$
by $360^{\circ}$ which leaves it invariant. Thus $\alpha^{5} = e$ which is the
first of the generator relations in Eq.~(\ref{dihedralrels}). Similarly, $\beta$
can be shown to correspond to a reflection of $C_{5}$ about a vertical axis 
passing through vertex $0$ in Figure~\ref{fig13}. Thus two applications of 
$\beta$ leave $C_{5}$ invariant. Thus $\beta^{2} =e$ which is the second
generator relation in Eq.~(\ref{dihedralrels}). Finally, using 
Table~\ref{table2}, direct calculation as in Eqs.~(\ref{testrel1}) and 
(\ref{testrel2}) shows that $\alpha\beta = \beta\alpha^{4}$ which establishes
the final generator relation in Eq.~(\ref{dihedralrels}). We see that $\alpha$ 
and $\beta$ generate $Aut(C_{5})$ and satisfy the generator relations for
$D_{5}$ and so generate a $10$ element group isomorphic to $D_{5}$. In summary, 
we have shown that the GI AQA found all ten graph automorphisms of $C_{5}$, 
and that the group formed from these automorphisms is isomorphic to the 
dihedral group $D_{5}$ which is the correct automorphism group for $C_{5}$. \\

\begin{figure}[h!]
\begin{center}
\setlength{\unitlength}{0.1in}
\begin{picture}(30,20)(0,-2)
\thicklines
\put(10,12.5){\circle*{1}}
\put(10,2.5){\circle*{1}}
\put(20,12.5){\circle*{1}}
\put(20,2.5){\circle*{1}}
\put(5,7.5){\circle*{1}}
\put(25,7.5){\circle*{1}}

\put(10,2.5){\line(1,0){10}}
\put(10,12.5){\line(1,0){10}}
\put(5,7.5){\line(1,1){5}}
\put(5,7.5){\line(1,-1){5}}
\put(25,7.5){\line(-1,1){5}}
\put(25,7.5){\line(-1,-1){5}}

\put(8.5,13){$0$}
\put(8.5,1){$4$}
\put(20.7,13){$1$}
\put(20.7,1){$3$}
\put(25.7,8){$2$}
\put(3.5,8){$5$}
\put(14.5,-2){$\mathbf{\underline{C_{6}}}$}
\end{picture}
\end{center}
\caption{\label{fig14}Cycle graph $C_{6}$.}
\end{figure}
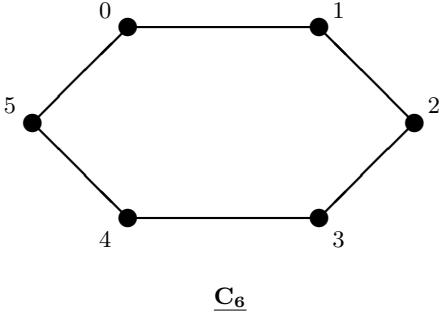

\underline{$N = 6$}: The cycle graph $C_{6}$ appears in Figure~\ref{fig14}.
It has $6$ vertices and $6$ edges; degree sequence $\{ 2,2,2,2,2,2\}$; and 
adjacency matrix
\begin{equation}
A = \left(  \begin{array}{cccccc}
                     0 & 1 & 0 & 0 & 0 & 1 \\
                     1 & 0 & 1 & 0 & 0 & 0 \\
                     0 & 1 & 0 & 1 & 0 & 0 \\
                     0 & 0 & 1 & 0 & 1 & 0 \\
                     0 & 0 & 0 & 1 & 0 & 1 \\
                     1 & 0 & 0 & 0 & 1 & 0 \\
                \end{array}
      \right) .
\label{C6admat}
\end{equation}

Numerical simulation of the GI AQA applied to the GI instance with $G = C_{6}$
and $G^{\prime} = G$ found a vanishing final ground-state energy $E_{gs} 
= 0$. The GI AQA thus correctly identifies $C_{6}$ as being isomorphic 
to itself. The simulation also found that the final ground-state subspace is
$12$-fold degenerate. Table~\ref{table3}
\begin{table}[h!]
\caption{\label{table3}Automorphism group $Aut(C_{6})$ of the cycle graph 
$C_{6}$ as found by the GI AQA. The first and third rows list the integer 
strings $s=s_{0}\cdots s_{5}$ determined by the labels of the twelve CBS 
$|s_{b}\rangle$ that span the final ground-state subspace (see text). Each 
string $s$ determines a graph automorphism $\pi (s)$ via 
Eq.~(\ref{permdefsec4c}). The second and fourth rows associate each integer 
string $s$ in the first and third rows with a graph automorphism $\pi (s)$. They
also identify the two strings that give rise to the graph automorphisms 
$\alpha$ and $\beta$ that generate $Aut(C_{6})$, and write each graph 
automorphism $\pi (s)$ as a product of an appropriate power of $\alpha$ 
and $\beta$. Note that $e$ is the identity automorphism, and the product 
notation assumes the rightmost factor acts first.\\}
\begin{ruledtabular}
\begin{tabular}{c|cccccc}
$s = s_{0}\cdots s_{5}$ & $501234$  &  $450123$ & $345012$ &  $234501$ &  
   $123450$  &  $012345$ \\ 
$\pi (s)$ & $\alpha$ &  $\alpha^{2}$ &  $\alpha^{3}$ &  $\alpha^{4}$ & 
    $\alpha^{5}$  &  $\alpha^{6} = e$  \\ \hline
$s = s_{0}\cdots s_{5}$ & $105432$ & $210543$ & $321054$ & $432105$ &
     $543210$  &  $054321$ \\ 
$\pi (s)$ & $\beta$  &  $\alpha\beta$  &  $\alpha^{2}\beta$  &  
     $\alpha^{3}\beta$ &  $\alpha^{4}\beta$  &   $\alpha^{5}\beta$ \\
\end{tabular}
\end{ruledtabular}
\end{table}
lists the integer strings $s = s_{0}\cdots s_{5}$ that result from the 
twelve CBS $|s_{b}\rangle$ that span the final ground-state subspace (see
Sections~\ref{sec2b} and \ref{sec3}). Each integer string $s$ fixes the
bottom row of a permutation $\pi (s)$ (see eq.~(\ref{permdefsec4c})) which is
a graph automorphism of $C_{6}$. The demonstration of this is identical to the 
demonstration given for $C_{4}$ and so will not be repeated here. Just as for
$Aut(C_{4})$, the graph automorphisms in Table~\ref{table3} are all the elements
of $Aut(C_{6})$ which is seen to have order $12$. Note that this is the same as
the order of  the dihedral group $D_{6}$.

We now show that $Aut(C_{6})$ is isomorphic to the dihedral group $D_{6}$ by
showing that the graph automorphisms $\alpha = \pi (501234)$ and $\beta =
\pi (105432)$ generate $Aut(C_{6})$, and satisfy the generator relations
(Eq.~(\ref{dihedralrels})) for $D_{6}$. The second and fourth rows of 
Table~\ref{table3} establish that $\alpha$ and $\beta$ are the generators 
of $Aut(C_{6})$ as they show that each element of $Aut(C_{6})$ is a product 
of an appropriate power of $\alpha$ and $\beta$, and that all possible products 
of powers of $\alpha$ and $\beta$ appear in these two rows. Following the 
discussion for $C_{4}$, it is a simple matter to show that $\alpha$ corresponds 
to a $60^{\circ}$ clockwise rotation of $C_{6}$. Thus $6$ applications of 
$\alpha$ rotates $C_{6}$ by $360^{\circ}$ which leaves it invariant. Thus 
$\alpha^{6} = e$ which is the first of the generator relations in 
Eq.~(\ref{dihedralrels}). Similarly, $\beta$ can be shown to correspond to a 
reflection of $C_{6}$ about a vertical axis that bisects $C_{6}$. Thus two 
applications of $\beta$ leave $C_{6}$ invariant. Thus $\beta^{2} =e$ which 
is the second generator relation in Eq.~(\ref{dihedralrels}). Finally, using 
Table~\ref{table3}, direct calculation as in Eqs.~(\ref{testrel1}) and 
(\ref{testrel2}) shows that $\alpha\beta = \beta\alpha^{5}$ which establishes
the final generator relation in Eq.~(\ref{dihedralrels}). We see that $\alpha$
and $\beta$ generate $Aut(C_{6})$ and satisfy the generator relations for
$D_{6}$ and so generate a $12$ element group isomorphic to $D_{6}$. In summary,
we have shown that the GI AQA found all twelve graph automorphisms of $C_{6}$,
and that the group formed from these automorphisms is isomorphic to the 
dihedral group $D_{6}$ which is the correct automorphism group for $C_{6}$. \\

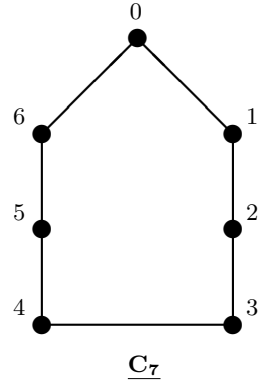
\begin{figure}[h!]
\begin{center}
\setlength{\unitlength}{0.1in}
\begin{picture}(30,18)(0,-5)
\thicklines
\put(10,7.5){\circle*{1}}
\put(10,2.5){\circle*{1}}
\put(20,7.5){\circle*{1}}
\put(20,2.5){\circle*{1}}
\put(15,12.5){\circle*{1}}
\put(10,-2.5){\circle*{1}}
\put(20,-2.5){\circle*{1}}

\put(10,-2.5){\line(0,1){10}}
\put(20,-2.5){\line(0,1){10}}
\put(10,7.5){\line(1,1){5}}
\put(20,7.5){\line(-1,1){5}}
\put(10,-2.5){\line(1,0){10}}

\put(14.6,13.5){$0$}
\put(8.5,3){$5$}
\put(20.7,8){$1$}
\put(20.7,3){$2$}
\put(8.5,8){$6$}
\put(8.5,-2){$4$}
\put(20.7,-2){$3$}

\put(14.5,-5){$\mathbf{\underline{C_{7}}}$}
\end{picture}
\end{center}
\caption{\label{fig15}Cycle graph $C_{7}$.}
\end{figure}

\underline{$N = 7$}: The cycle graph $C_{7}$ appears in Figure~\ref{fig15}.
It has $7$ vertices and $7$ edges; degree sequence $\{ 2,2,2,2,2,2,2\}$; and
adjacency matrix
\begin{equation}
A = \left(  \begin{array}{ccccccc}
                     0 & 1 & 0 & 0 & 0 & 0 & 1 \\
                     1 & 0 & 1 & 0 & 0 & 0 & 0 \\
                     0 & 1 & 0 & 1 & 0 & 0 & 0 \\
                     0 & 0 & 1 & 0 & 1 & 0 & 0 \\
                     0 & 0 & 0 & 1 & 0 & 1 & 0 \\
                     0 & 0 & 0 & 0 & 1 & 0 & 1 \\
                     1 & 0 & 0 & 0 & 0 & 1 & 0 \\
                \end{array}
      \right) .
\label{C7admat}
\end{equation}

Numerical simulation of the GI AQA applied to the GI instance with $G = C_{7}$
and $G^{\prime} = G$ found a vanishing final ground-state energy $E_{gs} 
= 0$. The GI AQA thus correctly identifies $C_{7}$ as being isomorphic 
to itself. The simulation also found that the final ground-state subspace is
$14$-fold degenerate. Table~\ref{table4}
\begin{table}[h!]
\caption{\label{table4}Automorphism group $Aut(C_{7})$ of the cycle graph 
$C_{7}$ as found by the GI AQA. The odd rows list the integer 
strings $s=s_{0}\cdots s_{6}$ determined by the labels of the fourteen CBS 
$|s_{b}\rangle$ that span the final ground-state subspace (see text). Each 
string $s$ determines a graph automorphism $\pi (s)$ via 
Eq.~(\ref{permdefsec4c}). Each even row associates each integer 
string $s$ in the odd row preceding it with a graph automorphism $\pi (s)$. 
They also identify the two strings that give rise to the graph automorphisms 
$\alpha$ and $\beta$ that generate $Aut(C_{7})$, and write each graph 
automorphism $\pi (s)$ as a product of an appropriate power of $\alpha$ 
and $\beta$. Note that $e$ is the identity automorphism, and the product 
notation assumes the rightmost factor acts first.\\}
\begin{ruledtabular}
\begin{tabular}{c|ccccccc}
$s = s_{0}\cdots s_{6}$ & $6012345$  &  $5601234$ & $4560123$ &  $3456012$ \\
$\pi (s)$ & $\alpha$ &  $\alpha^{2}$ &  $\alpha^{3}$ &  $\alpha^{4}$  \\  \hline

$s = s_{0}\cdots s_{6}$ &  $2345601$  &  $1234560$  & $0123456$ \\ 
$\pi (s)$  & $\alpha^{5}$  &  $\alpha^{6}$ & $\alpha^{7} = e$  \\ \hline

$s = s_{0}\cdots s_{6}$ & $0654321$ & $1065432$ & $2106543$ & $3210654$ \\
$\pi (s)$ & $\beta$  &  $\alpha\beta$  &  $\alpha^{2}\beta$  &  
     $\alpha^{3}\beta$ \\\hline

$s = s_{0}\cdots s_{6}$  & $4321065$  &  $5432106$ &  $6543210$ \\ 
$\pi (s)$  &  $\alpha^{4}\beta$  &   $\alpha^{5}\beta$ & 
         $\alpha^{6}\beta$ \\
\end{tabular}
\end{ruledtabular}
\end{table}
lists the integer strings $s = s_{0}\cdots s_{6}$ that result from the 
fourteen CBS $|s_{b}\rangle$ that span the final ground-state subspace (see
Sections~\ref{sec2b} and \ref{sec3}). Each integer string $s$ fixes the
bottom row of a permutation $\pi (s)$ (see eq.~(\ref{permdefsec4c})) which is
a graph automorphism of $C_{7}$. The demonstration of this is identical to the 
demonstration given for $C_{4}$ and so will not be repeated here. Just as for
$Aut(C_{4})$, the graph automorphisms in Table~\ref{table4} are all the elements
of $Aut(C_{7})$ which is seen to have order $14$. Note that this is the same as
the order of  the dihedral group $D_{7}$.

We now show that $Aut(C_{7})$ is isomorphic to the dihedral group $D_{7}$ by
showing that the graph automorphisms $\alpha = \pi (6012345)$ and $\beta =
\pi (0654321)$ generate $Aut(C_{7})$, and satisfy the generator relations
(Eq.~(\ref{dihedralrels})) for $D_{7}$. The second and fourth rows of 
Table~\ref{table4} establish that $\alpha$ and $\beta$ are the generators 
of $Aut(C_{7})$ as they show that each element of $Aut(C_{7})$ is a product 
of an appropriate power of $\alpha$ and $\beta$, and that all possible products 
of powers of $\alpha$ and $\beta$ appear in these two rows. Following the 
discussion for $C_{4}$, it is a simple matter to show that $\alpha$ corresponds 
to a $2\pi /7$ radian clockwise rotation of $C_{7}$. Thus $7$ applications of 
$\alpha$ rotates $C_{7}$ by $360^{\circ}$ which leaves it invariant. Thus 
$\alpha^{7} = e$ which is the first of the generator relations in 
Eq.~(\ref{dihedralrels}). Similarly, $\beta$ can be shown to correspond to a 
reflection of $C_{7}$ about a vertical axis that passes through vertex $0$ 
in Figure~\ref{fig15}. Thus two applications of $\beta$ leave $C_{7}$ invariant.
Thus $\beta^{2} =e$ which is the second generator relation in 
Eq.~(\ref{dihedralrels}). Finally, using Table~\ref{table4}, direct calculation 
as in Eqs.~(\ref{testrel1}) and (\ref{testrel2}) shows that $\alpha\beta = 
\beta\alpha^{6}$ which establishes the final generator relation in 
Eq.~(\ref{dihedralrels}). We see that $\alpha$ and $\beta$ generate $Aut(C_{7})$
and satisfy the generator relations for $D_{7}$ and so generate a $14$ element 
group isomorphic to $D_{7}$. In summary, we have shown that the GI AQA found 
all fourteen graph automorphisms of $C_{7}$, and that the group formed from 
these automorphisms is isomorphic to the dihedral group $D_{7}$ which is the 
correct automorphism group for $C_{7}$. \\

\subsubsection{Grid graph $G_{2,3}$}
\label{sec4c2}

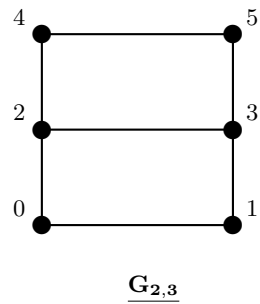
\begin{figure}[h!]
\begin{center}
\setlength{\unitlength}{0.1in}
\begin{picture}(30,17)(0,-1)
\thicklines
\put(10,12.5){\circle*{1}}
\put(10,2.5){\circle*{1}}
\put(20,12.5){\circle*{1}}
\put(20,2.5){\circle*{1}}
\put(10,7.5){\circle*{1}}
\put(20,7.5){\circle*{1}}

\put(10,2.5){\line(0,1){10}}
\put(20,2.5){\line(0,1){10}}
\put(10,2.5){\line(1,0){10}}
\put(10,12.5){\line(1,0){10}}
\put(10,7.5){\line(1,0){10}}
\put(8.5,13){$4$}
\put(8.5,3){$0$}
\put(20.7,13){$5$}
\put(20.7,3){$1$}
\put(8.5,8){$2$}
\put(20.7,8){$3$}

\put(14.5,-1){$\mathbf{\underline{G_{2,3}}}$}
\end{picture}
\end{center}
\caption{\label{fig16}Grid graph $G_{2,3}$.}
\end{figure}

The grid graph $G_{2,3}$ appears in Figure~\ref{fig16}.
It has $6$ vertices and $7$ edges; degree sequence $\{ 3,3,2,2,2,2\}$; and
adjacency matrix
\begin{equation}
A = \left(  \begin{array}{cccccc}
                     0 & 1 & 1 & 0 & 0 & 0 \\
                     1 & 0 & 0 & 1 & 0 & 0 \\
                     1 & 0 & 0 & 1 & 1 & 0 \\
                     0 & 1 & 1 & 0 & 0 & 1 \\
                     0 & 0 & 1 & 0 & 0 & 1 \\
                     0 & 0 & 0 & 1 & 1 & 0 \\
                \end{array}
      \right) .
\label{G23admat}
\end{equation}
$G_{2,3}$ has two reflection symmetries. The first is reflection about the
horizontal line passing through vertices $2$ and $3$, and the second is 
reflection about a vertical line that bisects $G_{2,3}$. Let $\alpha$ and 
$\beta$ denote the permutation of vertices that these reflections produce. 
It follows from their definition that $\alpha = \pi (452301)$ and $\beta = 
\pi (103254)$. They are the generators of the automorphism group of 
$G_{2,3}$: $Aut(G_{2,3}) = \langle \alpha ,\beta\rangle$. As reflections, 
they satisfy $\alpha^{2}=\beta^{2} = e$. Applying these reflections 
successively gives $\alpha\beta = \pi (543210)$ which is a fourth symmetry 
of $G_{2,3}$. Applying these reflections in the reverse order, it is easy to 
verify that $\alpha\beta = \beta\alpha$. Thus $Aut(G_{2,3})$ is a $4$ element 
Abelian group generated by the reflections $\alpha$ and $\beta$. Let us now
compare this with the results found by the GI AQA.

Numerical simulation of the GI AQA applied to the GI instance with $G = G_{2,3}$
and $G^{\prime} = G$ found a vanishing final ground-state energy $E_{gs} 
= 0$. The GI AQA thus correctly identifies $G_{2,3}$ as being isomorphic 
to itself. The simulation also found that the final ground-state subspace is
$4$-fold degenerate. Table~\ref{table5}
\begin{table}[h!]
\caption{\label{table5}Automorphism group $Aut(G_{2,3})$ of the grid graph
$G_{2,3}$ as found by the GI AQA. The first row lists the integer strings 
$s=s_{0}\cdots s_{5}$ determined by the labels of the four CBS 
$|s_{b}\rangle$ that span the final ground-state subspace (see text). Each 
string $s$ determines a graph automorphism $\pi (s)$ via 
Eq.~(\ref{permdefsec4c}). The second row associates each integer 
string $s$ in the first row with a graph automorphism $\pi (s)$. They
also identify the two strings that give rise to the graph automorphisms 
$\alpha$ and $\beta$ that generate $Aut(G_{2,3})$, and write each graph 
automorphism $\pi (s)$ as a product of an appropriate power of $\alpha$ 
and $\beta$. Note that $e$ is the identity automorphism, and the product 
notation assumes the rightmost factor acts first.\\}
\begin{ruledtabular}
\begin{tabular}{c|cccc}
$s = s_{0}\cdots s_{5}$ & $452301$  &  $103254$ & $012345$ & 
     $543210$ \\ \hline
$\pi (s)$ & $\alpha$ &  $\beta$ &  $\alpha^{2}=\beta^{2} = e$ &  
   $\alpha\beta$ \\ 
\end{tabular}
\end{ruledtabular}
\end{table}
lists the integer strings $s = s_{0}\cdots s_{5}$ that result from the 
four CBS $|s_{b}\rangle$ that span the final ground-state subspace (see
Sections~\ref{sec2b} and \ref{sec3}). Each integer string $s$ fixes the
bottom row of a permutation $\pi (s)$ (see eq.~(\ref{permdefsec4c})) which is
a graph automorphism of $G_{2,3}$. The demonstration of this is identical to the 
demonstration given for $C_{4}$ and so will not be repeated here. Notice that 
the GI AQA found the graph automorphisms $\pi (452301)$ and $\pi (103254)$ 
which implement the reflection symmetries of $G_{2,3}$ described above. As 
they implement reflections, it follows that $\alpha^{2} = \beta^{2} = e$. The
GI AQA also found the graph automorphism $\pi (543210)$ which is the composite
symmetry $\alpha\beta$ described above. Using Table~\ref{table5}, direct 
calculation as in Eqs.~(\ref{testrel1}) and (\ref{testrel2}) shows that 
$\alpha\beta = \beta\alpha$. Thus the GI AQA correctly found the two generators 
$\alpha$ and $\beta$ of $Aut(G_{2,3})$; correctly determined all 4 elements of 
$Aut(G_{2,3})$; and correctly determined that $Aut(G_{2,3})$ is an Abelian 
group. In summary, the GI AQA correctly determined the automorphism group of 
$G_{2,3}$. \\

\subsubsection{Wheel graph $W_{7}$}
\label{sec4c3}

The wheel graph $W_{7}$ appears in Figure~\ref{fig17}.
\begin{figure}[h!]
\begin{center}
\setlength{\unitlength}{0.1in}
\begin{picture}(30,20)(0,-5)
\thicklines
\put(10,12.5){\circle*{1}}
\put(10,2.5){\circle*{1}}
\put(20,12.5){\circle*{1}}
\put(20,2.5){\circle*{1}}
\put(5,7.5){\circle*{1}}
\put(25,7.5){\circle*{1}}
\put(15,7.5){\circle*{1}}

\put(10,2.5){\line(1,1){10}}
\put(10,12.5){\line(1,-1){10}}
\put(10,2.5){\line(1,0){10}}
\put(10,12.5){\line(1,0){10}}
\put(5,7.5){\line(1,1){5}}
\put(5,7.5){\line(1,-1){5}}
\put(25,7.5){\line(-1,1){5}}
\put(25,7.5){\line(-1,-1){5}}
\put(5,7.5){\line(1,0){20}}

\put(8.5,13){$0$}
\put(8.5,1){$4$}
\put(20.7,13){$1$}
\put(20.7,1){$3$}
\put(25.7,8){$2$}
\put(3.5,8){$5$}
\put(14.7,8.5){$6$}

\put(14.5,-5){$\mathbf{\underline{W_{7}}}$}
\end{picture}
\end{center}
\caption{\label{fig17}Wheel graph $W_{7}$.}
\end{figure}
It has $7$ vertices and $12$ edges; degree sequence $\{ 6,3,3,3,3,3,3\}$; and
adjacency matrix
\begin{equation}
A = \left(  \begin{array}{ccccccc}
                     0 & 1 & 0 & 0 & 0 & 1 & 1 \\
                     1 & 0 & 1 & 0 & 0 & 0 & 1 \\
                     0 & 1 & 0 & 1 & 0 & 0 & 1 \\
                     0 & 0 & 1 & 0 & 1 & 0 & 1 \\
                     0 & 0 & 0 & 1 & 0 & 1 & 1 \\
                     1 & 0 & 0 & 0 & 1 & 0 & 1 \\
                     1 & 1 & 1 & 1 & 1 & 1 & 0 \\
                \end{array}
      \right) .
\label{W7admat}
\end{equation}
The automorphism group $Aut(W_{7})$ is isomorphic to the dihedral group $D_{6}$
which, as we have seen, has $12$ elements and is generated by two graph 
automorphisms $\alpha$ and $\beta$ that satisfy the generator relations in 
Eq.~(\ref{dihedralrels}) with $n=6$ and respectively, rotate $W_{7}$ 
clockwise about vertex $6$ by $60^{\circ}$, and reflect $W_{7}$ about a 
vertical axis through vertex $6$. The graph automorphisms of $W_{7}$ 
thus fix vertex $6$. Let us now compare this with the results found by 
the GI AQA.

Numerical simulation of the GI AQA applied to the GI instance with $G = W_{7}$
and $G^{\prime} = G$ found a vanishing final ground-state energy $E_{gs} 
= 0$. The GI AQA thus correctly identifies $W_{7}$ as being isomorphic 
to itself. The simulation also found that the final ground-state subspace is
$12$-fold degenerate. Table~\ref{table6}
\begin{table}[h!]
\caption{\label{table6}Automorphism group $Aut(W_{7})$ of the wheel graph 
$W_{7}$ as found by the GI AQA. The odd rows list the integer 
strings $s=s_{0}\cdots s_{6}$ determined by the labels of the twelve CBS 
$|s_{b}\rangle$ that span the final ground-state subspace (see text). Each 
string $s$ determines a graph automorphism $\pi (s)$ via 
Eq.~(\ref{permdefsec4c}). Each even row associates each integer 
string $s$ in the odd row preceding it with a graph automorphism $\pi (s)$. 
They also identify the two strings that give rise to the graph automorphisms 
$\alpha$ and $\beta$ that generate $Aut(W_{7})$, and write each graph 
automorphism $\pi (s)$ as a product of an appropriate power of $\alpha$ 
and $\beta$. Note that $e$ is the identity automorphism, and the product 
notation assumes the rightmost factor acts first.\\}
\begin{ruledtabular}
\begin{tabular}{c|cccccc}
$s = s_{0}\cdots s_{6}$ & $5012346$  &  $4501236$ & $3450126$ \\
$\pi (s)$ & $\alpha$ &  $\alpha^{2}$ &  $\alpha^{3}$ \\ \hline
  
$s = s_{0}\cdots s_{6}$ &  $2345016$ &  $1234506$  &  $0123456$  \\ 
 $\pi (s)$  &  $\alpha^{4}$  & $\alpha^{5}$  &  $\alpha^{6} = e$  \\ \hline

$s = s_{0}\cdots s_{6}$ & $1054326$ & $2105436$ & $3210546$  \\
$\pi (s)$  & $\beta$  &  $\alpha\beta$  &  $\alpha^{2}\beta$   \\ \hline

$s = s_{0}\cdots s_{6}$ & $4321056$   & $5432106$  &  $0543216$ \\ 
$\pi (s)$ & $\alpha^{3}\beta$ &  $\alpha^{4}\beta$  &   $\alpha^{5}\beta$ \\
\end{tabular}
\end{ruledtabular}
\end{table}
lists the integer strings $s = s_{0}\cdots s_{6}$ that result from the 
twelve CBS $|s_{b}\rangle$ that span the final ground-state subspace (see
Sections~\ref{sec2b} and \ref{sec3}). Each integer string $s$ fixes the
bottom row of a permutation $\pi (s)$ (see eq.~(\ref{permdefsec4c})) which is
a graph automorphism of $W_{7}$. The demonstration of this is identical to the 
demonstration given for $C_{4}$ and so will not be repeated here. Just as for
$Aut(C_{4})$, the graph automorphisms in Table~\ref{table6} are all the elements
of $Aut(W_{7})$ which is seen to have order $12$. Note that this is the same as
the order of  the dihedral group $D_{6}$.

We now show that $Aut(W_{7})$ is isomorphic to the dihedral group $D_{6}$ by
showing that the graph automorphisms $\alpha = \pi (5012346)$ and $\beta =
\pi (1054326)$ generate $Aut(W_{7})$, and satisfy the generator relations
(Eq.~(\ref{dihedralrels})) for $D_{6}$. The second and fourth rows of 
Table~\ref{table6} establish that $\alpha$ and $\beta$ are the generators 
of $Aut(W_{7})$ as they show that each element of $Aut(W_{7})$ is a product 
of an appropriate power of $\alpha$ and $\beta$, and that all possible products 
of powers of $\alpha$ and $\beta$ appear in these two rows. Following the 
discussion for $C_{4}$, it is a simple matter to show that $\alpha$ corresponds 
to a $60^{\circ}$ clockwise rotation about vertex $6$ of $W_{7}$. Thus $6$ 
applications of $\alpha$ rotates $W_{7}$ by $360^{\circ}$ which leaves it 
invariant. Thus $\alpha^{6} = e$ which is the first of the generator relations in 
Eq.~(\ref{dihedralrels}). Similarly, $\beta$ can be shown to correspond to a 
reflection of $W_{7}$ about a vertical axis passing through vertex $6$ of 
$W_{7}$. Thus two applications of $\beta$ leave $W_{7}$ invariant. Thus 
$\beta^{2} =e$ which is the second generator relation in 
Eq.~(\ref{dihedralrels}). Finally, using Table~\ref{table6}, direct calculation 
as in Eqs.~(\ref{testrel1}) and (\ref{testrel2}) shows that $\alpha\beta = 
\beta\alpha^{5}$ which establishes the final generator relation in 
Eq.~(\ref{dihedralrels}). We see that $\alpha$ and $\beta$ generate 
$Aut(W_{7})$ and satisfy the generator relations for $D_{6}$ and so generate 
a $12$ element group isomorphic to $D_{6}$. In summary, we have shown that 
the GI AQA found all twelve graph automorphisms of $W_{7}$, and that the 
group formed from these automorphisms is isomorphic to the dihedral group 
$D_{6}$ which is the correct automorphism group for $W_{7}$.

\section{Experimental implementation}
\label{sec5}

In this section we express the GI problem Hamiltonian $H_{P}$ in a form
more suitable for experimental implementation. We saw in 
Section~\ref{sec2c} that the eigenvalues of $H_{P}$ are given by the cost 
function $C(s)$ which is reproduced here for convenience:
\begin{equation}
C(s) = C_{1}(s) + C_{2}(s) + C_{3}(s) ,
\label{costfuncdefv2}
\end{equation}
with 
\begin{eqnarray}
C_{1}(s) & = & \sum_{i=0}^{N-1}\sum_{\alpha = N}^{M} \delta_{s_{i},\alpha} 
                                        \label{C1def}\\
C_{2}(s) & = & \sum_{i=0}^{N-2}\sum_{j = i+1}^{N-1} \delta_{s_{i},s_{j}} ,
                               \label{C2def}
\end{eqnarray}
\begin{equation}
C_{3}(s) = \| \sigma (s)A\sigma^{T}(s) - A^{\prime}\|_{i} .
\label{C3def}
\end{equation}
Here $s = s_{0}\cdots s_{N-1}$ is the integer string derived from the binary
string $s_{b} = (s_{0}\cdots s_{U-1})\cdots (s_{(N-1)U}\cdots s_{NU-1})$
via Eqs.~(\ref{parsestr}) and (\ref{substr2int}) with $U = \lceil \log_{2}N
\rceil$. The matrix $\sigma (s)$ was defined in Eq.~(\ref{sigmatelm}) as 
\begin{equation}
\sigma_{i,j}(s) = 
\left\{ \begin{array}{cl}
               0,\hspace{0.2in} {} & \mathrm{if}\: s_{j} > N-1\\
               \delta_{i,s_{j}}, & \mathrm{if} \: 0 \leq s_{j} \leq N-1 .
           \end{array} \right.
\label{sigmatelmv2}
\end{equation}
Note that $\sigma_{i,j}(s)$ can be written more compactly as
\begin{equation}
\sigma_{i,j}(s) = \delta_{i,s_{j}}\prod_{\alpha = N}^{M}\left(
                                   1 - \delta_{s_{j},\alpha}\right) .
\label{bettersigmaij}
\end{equation}
We see from Eqs.~(\ref{costfuncdefv2})--(\ref{C3def}) and 
(\ref{bettersigmaij}) that the $s$-dependence of $C(s)$ enters through the
Kronecker deltas. This type of $s$-dependence is not well-suited for 
experimental implementation and so our task is to find a more convenient
form for the Kronecker delta.

We begin with $\delta_{a,b}$ in the case where $a,b\in\{ 0,1\}$. Here we
write
\begin{equation}
\delta_{a,b} = \left( a + b -1\right)^{2} = 
    \left\{ \begin{array}{ccl}
                   0 & & (a\neq b) \\
                  1 & & (a=b) ,
              \end{array}
    \right.
\label{binaryKron}
\end{equation}
which can be checked by inserting values for $a$ and $b$. Now consider
$\delta_{s,k}$ when $s$ and $k$ are $U$-bit integers. The binary decompositions
of $s$ and $k$ are
\begin{eqnarray}
s & = & \sum_{i=0}^{U-1} s_{i}\left( 2\right)^{i} \\
k & = & \sum_{i=0}^{U-1} k_{i}\left( 2\right)^{i} .
\end{eqnarray}
For $s$ and $k$ to be equal, all their corresponding bits must be equal. Thus 
we can write
\begin{eqnarray}
\delta_{s,k} & = & \prod_{i=0}^{U-1}\delta_{s_{i},k_{i}} \nonumber \\
   & = &   \prod_{i=0}^{U-1}\left( s_{i}+k_{i}-1\right)^{2} =
    \left\{ \begin{array}{ccl}
                  1 & & (\mathrm{all\;}  s_{i}=k_{i}) \\
                  0 & & (  \mathrm{some\; } s_{i}\neq k_{i}). \\
              \end{array}  
   \right.   
\label{integerKron}
\end{eqnarray}

Eq.~(\ref{integerKron}) allows each Kronecker delta appearing in $C(s)$ to be
converted to a polynomial in the components of the integer string $s$. From
Eqs.~(\ref{C1def}) and (\ref{C2def}) we see that $C_{1}(s)$ and $C_{2}(s)$ 
are each $2U$-local. For $C_{3}(s)$ we must write $\sigma (s)A\sigma^{T}(s)$ 
in a form that makes the Kronecker deltas explicit. Using 
Eq.~(\ref{bettersigmaij}), we have
\begin{eqnarray}
\sigma (s)A\sigma^{T}(s) & = & \sum_{i,j=0}^{N-1} \sigma_{li}(s) A_{ij}
                                                       \sigma_{mj}(s) \nonumber\\
 & = & \sum_{i,j=0}^{N-1}\left\{ \delta_{l,s_{i}}\prod_{\alpha =N}^{M}
                 \left( 1 - \delta_{s_{i},\alpha}\right)\right\} A_{ij} \nonumber \\
  & & {}\hspace{0.1in}\times 
           \left\{ \delta_{m,s_{j}}\prod_{\beta =N}^{M}\left( 1 -
                     \delta_{s_{j},\beta}\right)\right\} . \label{sAsmatelm}
\end{eqnarray}
Inserting Eq.~(\ref{integerKron}) into Eq.~(\ref{sAsmatelm}), we see that
$\sigma (s)A\sigma^{T}(s) - A^{\prime}$ is $4U(M-N+1)$-local. If we use the 
$L_{1}$-norm ($L_{2}$-norm) in Eq.~(\ref{C3def}), then we see that $C_{3}(s)$
is $4U(M-N+1)$-local ($8U(M-N+1)$-local). 

The number of terms in $C(s)$, and thus in $H_{P}$, follows straight-forwardly 
from Eqs.~(\ref{C1def})-(\ref{C3def}). Starting with $C_{1}(s)$, there is a term
for each value of $i$ and $\alpha$ that appears in the sum. There are thus 
$T_{1} = N(M-N+1)$ terms in $C_{1}(s)$. From Eq.~(\ref{Mdef}), $M = 2^{U}-1$,
where $U = \lceil \log_{2} N\rceil$ and $N$ is the number of vertices in each 
graph appearing in the GI instance. It follows from the definition of $U$ that
$M < 2N$ and so $T_{1} < N(N+1)$. Recall from Section~\ref{sec3} that the 
number of qubits $L_{N}$ needed for an $N$-vertex GI instance is $L_{N} = 
N\lceil \log_{2} N\rceil$. Thus $T_{1} < (L_{N}/\lceil \log_{2} N\rceil )
(L_{N}/\lceil\log_{2}N\rceil + 1) < CL_{N}^{2}$, where $C$ is an appropriate 
constant. Thus $T_{1}(L_{N}) = \mathcal{O}(L_{N}^{2})$. A similar analysis 
shows that the number of terms in $C_{2}(s)$ is $T_{2}(L_{N}) = \mathcal{O}
(L_{N}^{2})$. Finally, $C_{3}(s)$ is the $L_{i}$-norm of an $N\times N$ matrix.
For the $L_{1}$-norm used in our simulations, the number of terms in $C_{3}(s)$
is $T_{3} = N^{2}$. Following the above analysis, this gives $T_{3}(L_{N}) = 
\mathcal{O}(L_{N}^{2})$. Putting everything together gives that the total 
number of terms associated with $C(s)$, and so also $H_{P}$, is $T(L_{N}) = 
T_{1}(L_{N})+T_{2}(L_{N})+T_{3}(L_{N}) = \mathcal{O}(L_{N}^{2})$. The initial 
Hamiltonian $H_{i}$ contains one term for each qubit (see Eq.~(\ref{HIdef})) 
and so the number of terms in $H_{i}$ is $L_{N}$. Thus the total number of 
terms in the full time-dependent Hamiltonian H(t) is $\mathcal{O}(L_{N}^{2})$ 
and so scales quadratically with the number of qubits $L_{N}$.

Clearly, the degree of difficulty associated with experimentally implementing 
a quantum algorithm depends strongly on the architecture of the hardware 
on which it is to be run. The simplest situation would be an architecture which
allows an arbitrary number of qubits to be simultaneously coupled, independently
of where they were located on the processor. Unfortunately, such a hardware 
architecture does not presently exist. For hardware platforms designed 
to run adiabatic quantum optimization algorithms, the D-Wave hardware is 
furthest along \cite{Johnson}. Each qubit on the D-Wave processor couples to at
most six neighboring qubits, and only two-qubit Ising coupling interactions 
are possible. The initial Hamiltonian $H_{i}$ (see Eq.~(\ref{HIdef})) is 
easily programmed onto the hardware. However, a problem Hamiltonian $H_{P}$ 
which is not of Ising form is more challenging, requiring an embedding 
procedure that: (i)~reduces all $k$-local interactions with $k\geq 3$ 
to $2$-local form; and (ii)~two-qubit coupling interactions that match the 
hardware's Chimera coupling-graph. A procedure for carrying out 
this reduction based on Ref.~\cite{bian} is described in 
Appendix~\ref{appendixEmBed}. Alternative approaches appear in 
Refs.~\cite{Babb} and \cite{Whit}.

\section{SubGraph Isomorphism problem}
\label{sec6}

An instance of the SubGraph Isomorphism (SGI) problem consists of an $N$-vertex
graph $G$ and an $n$-vertex graph $H$ with $n\leq N$. The question to be 
answered is whether $G$ contains an $n$-vertex subgraph that is isomorphic to 
$H$. The SGI problem is known to be NP-Complete \cite{Garey&Johnson} and
is believed to be more difficult to solve than the GI problem. Here we show 
how an instance of SGI can be converted into an instance of a COP whose cost 
function has a minimum value that vanishes  when $G$ contains a subgraph 
isomorphic to $H$, and is greater than zero otherwise. The SGI cost function 
will be seen to be a natural generalization of the GI cost function given in 
Eqs.~(\ref{costfuncdef})--(\ref{C3defv0}). The SGI COP can then be solved using
adiabatic quantum evolution as was done for the GI problem.

As with the GI problem, we would like to determine whether there exists an
isomorphism $\pi$ of $G$ that produces a new graph $\pi (G)$ that contains 
$H$ as a subgraph. Just as with the GI problem, we consider linear maps 
$\sigma (s)$ (see Eqs.~(\ref{parsestr})--(\ref{sigmatelm})) that transform 
the adjacency matrix $A$ of $G$ to $\tilde{A}(s) =\sigma (s)A\sigma^{T}(s)$. 
We then search $\pi (G)$ to determine whether there is a subset of $n$ 
vertices that yields a subgraph that is equal to $H$. 

To begin the process of converting an SGI instance into an instance of a COP,
let: (i)~$\alpha$ label all the $\binom{N}{n}$ ways of
choosing $n$ vertices from the $N$ vertices in $\pi (G)$; and (ii)~$|i\rangle$ 
($|\alpha_{i}\rangle$) be an $n$-component ($N$-component) vector whose
$i$-th ($\alpha_{i}$-th) component is $1$, and all other components are $0$. 
Thus $\alpha$ labels the choice $(\alpha_{0}, \ldots ,\alpha_{n-1})$ of
$n$ vertices out of the $N$ vertices of $\pi (G)$. We now show that an 
$n\times N$ matrix $P_{\alpha}$ can be used to form an $n\times n$ matrix
$\calA_{\alpha}(s)$ whose matrix elements are the matrix elements of 
$\tilde{A}(s)$ associated with the $n$ vertices appearing in $\alpha$. To that
purpose, define
\begin{equation}
P_{\alpha} = \sum_{i=0}^{n-1} |i\rangle\langle\alpha_{i}| ,
\label{Palphadef}
\end{equation}
\begin{equation}
\calA_{\alpha}(s) = P_{\alpha} \tilde{A}(s)P_{\alpha}^{T} ,
\label{Aalphadef}
\end{equation}
where 
\begin{equation}
\tilde{A}(s) = \sum_{l,m=0}^{N-1} \tilde{A}_{l,m}(s)|l\rangle\langle m|.
\end{equation}
It follows from these definitions that
\begin{eqnarray}
\calA_{\alpha}(s) & = & \sum_{i,=0}^{n-1}|i\rangle\langle\alpha_{i}| 
           \sum_{l,m=0}^{N-1}\tilde{A}_{l,m}(s)|l\rangle\langle m|
            \sum_{j=0}^{n-1}|\alpha_{j}\rangle\langle j| \nonumber\\
 & = & \sum_{i,j=0}^{n-1} \tilde{A}_{\alpha_{i},\alpha_{j}}(s)
                   |i\rangle\langle j| .
\end{eqnarray}
Thus the matrix elements $(\calA_{\alpha})_{i,j}(s)$ are precisely the 
matrix elements $\tilde{A}_{\alpha_{i},\alpha_{j}}(s)$ associated
with all possible pairs of vertices drawn from $(\alpha_{0}, \ldots , 
\alpha_{n-1})$. The matrix $\calA_{\alpha}(s)$ is thus the adjacency matrix 
for the subgraph $g_{\alpha}$ composed of the vertices appearing in $\alpha$, 
along with all the edges in $\pi (G)$ that join them. With $\calA_{\alpha}(s)$,
we can, as in the GI problem, test whether $g_{\alpha}$ is equal to $H$ by 
checking whether $||\calA_{\alpha}(s) - A^{\prime}||_{i}$ vanishes or not. 
Here $A^{\prime}$ is the adjacency matrix of the graph $H$, and 
$|| \mathcal{O}||_{i}$ is the $L_{i}$-norm of $\mathcal{O}$.

We can now define a cost function whose minimum value vanishes 
if and only if $G$ contains a subgraph $g$ that is isomorphic to $H$. Because 
the transformation $\sigma (s)$ must be a permutation matrix when $g$ is 
isomorphic to $H$, we again introduce the penalty functions $C_{1}(s)$ and 
$C_{2}(s)$ used in the GI COP to penalize those integer strings $s$ that 
produce a $\sigma (s)$ that is not a permutation matrix:
\begin{eqnarray}
C_{1}(s) & = & \sum_{i=0}^{N-1}\sum_{\alpha = N}^{M} \delta_{s_{i},\alpha} 
                                        \label{C1defv2}\\
C_{2}(s) & = & \sum_{i=0}^{N-2}\sum_{j = i+1}^{N-1} \delta_{s_{i},s_{j}} .
                               \label{C2defv2}
\end{eqnarray}
The final penalty function $C_{3}(s)$ for the SGI problem generalizes the one
used in the GI problem. It is defined to be
\begin{equation}
C_{3}(s) = \prod_{\alpha =1}^{\binom{N}{n}} ||\calA_{\alpha} (s) - A^{\prime}
                           ||_{i} ,
\label{C3defv2}
\end{equation}
where the product is over all $\binom{N}{n}$ ways of choosing $n$ vertices out
of $N$ vertices. Note that $C_{3}(s)$ vanishes if and only if $G$ contains an
$n$-vertex subgraph isomorphic to $H$. This follows since, if $G$ contains an
$n$-vertex subgraph $g$ isomorphic to $H$, there exists a permutation 
$\pi (s)$ of $G$ that has an $n$-vertex subgraph that is equal to $H$. Thus, 
for this $s$, there is a choice $\alpha$ of $n$ vertices that gives a subgraph
for which $\calA_{\alpha}(s) - A^{\prime} = 0$. It follows from
Eq.~(\ref{C3defv2}) that $C_{3}(s) = 0$. On the other hand, if $C_{3}(s) 
= 0$, it follows that at least one of the factor on the RHS of 
Eq.~(\ref{C3defv2}) vanishes. Thus there is a choice $\alpha$ of $n$ vertices 
for which $||\calA_{\alpha} (s) - A^{\prime}||_{i} = 0$. Thus 
$\calA_{\alpha}(s) = A^{\prime}$, and so $G$ has a subgraph isomorphic to $H$.

The cost function for the SGI problem is now defined to be
\begin{equation}
C(s) = C_{1}(s) + C_{2}(s) + C_{3}(s),
\label{costfuncv2}
\end{equation}
where $C_{1}(s)$, $C_{2}(s)$, and $C_{3}(s)$ are defined in 
Eqs.~(\ref{C1defv2})--(\ref{C3defv2}). This gives rise to the following COP:
\begin{description}
\item[SubGraph Isomorphism COP:] Given an $N$-vertex graph $G$ and an
$n$-vertex graph $H$ with $n\leq N$, and the associated cost function $C(s)$ 
defined in Eq.~(\ref{costfuncv2}), find an integer string $s_{\ast}$ that 
minimizes $C(s)$.
\end{description}
By construction: (i)~$C(s_{\ast}) = 0$ if and only if $G$ contains a subgraph
isomorphic to $H$, and $\sigma (s_{\ast})$ is the permutation matrix that
transforms $G$ into a graph $\pi (G)$ that has $H$ as a subgraph; and 
(ii)~$C(s_{\ast}) > 0$ otherwise.

As with the GI COP, the SGI COP can be solved using adiabatic quantum evolution.
 The quantum algorithm for SGI begins by preparing the $L = N\lceil \log_{2} N
\rceil$ qubit register in the ground-state of $H_{i}$ (see Eq.~(\ref{HIdef}))
and then driving the qubit register dynamics using the time-dependent 
Hamiltonian $H(t) = (1-t/T)H_{i}+(t/T)H_{P}$. Here the problem Hamiltonian
$H_{P}$ is defined to be diagonal in the computational basis $|s_{b}\rangle$ 
and to have associated eigenvalues $C(s)$, where $s$ is found from $s_{b}$ 
according to Eqs.~(\ref{parsestr})--(\ref{substr2int}). At the end of the 
evolution the qubits are measured in the computational basis. The outcome is 
the bit string $\ssb^{\ast}$ so that the final state of the register is 
$|\ssb^{\ast}\rangle$ and its energy is $C(s^{\ast})$, where $s^{\ast}$ is 
the integer string derived from $\ssb^{\ast}$. In the adiabatic limit, 
$C(s^{\ast})$ will be the ground-state energy, and if $C(s^{\ast}) =0$ 
($\: > 0$) the algorithm concludes that $G$ contains (does not contain) a 
subgraph isomorphic to $H$. In the case where $G$ does contain a subgraph 
isomorphic to $H$, the algorithm also returns the permutation $\pi_{\ast} =
\pi (s^{\ast})$ that converts $G$ to the graph $\pi_{\ast}(G)$ that contains
$H$ as a subgraph. Note that any real application of AQO will only be 
approximately adiabatic. Thus the probability that the final energy 
$C(s^{\ast})$ will be the ground-state energy will be $1-\epsilon$. In this 
case the SGI quantum algorithm must be run $k\sim \mathcal{O}
(\ln (1-\delta )/\ln\epsilon )$ times so that, with probability $\delta > 
1-\epsilon$, at least one of the measurements will return the ground-state 
energy. We can make $\delta$ arbitrarily close to $1$ by choosing $k$ 
sufficiently large.\\

\section{Summary}
\label{sec7}

In this paper we have presented a quantum algorithm that solves arbitrary 
instances of the Graph Isomorphism problem and which provides a novel
approach for finding the automorphism group of a graph. We numerically 
simulated the algorithm's quantum dynamics and showed that it correctly: 
(i)~distinguished non-isomorphic graphs; (ii)~recognized isomorphic graphs; 
and (iii)~determined the automorphism group of a given graph. We also 
discussed the quantum algorithm's experimental implementation, and showed 
how it can be generalized to give a quantum algorithm that solves arbitrary
instances of the NP-Complete SubGraph Isomorphism problem. As explained
in Appendix~\ref{appendixQAT}, the minimum energy gap for an adiabatic 
quantum algorithm largely determines the algorithm's computational complexity.
Determining this gap in the limit of large problem size is currently an 
important open problem in adiabatic quantum computing (see Section~\ref{sec4}).
It is thus not possible to determine the computational complexity of adiabatic 
quantum algorithms in general, nor, consequently, of the specific adiabatic 
quantum algorithms presented in this paper. Adiabatic quantum computing has
been shown to be equivalent to the circuit-model of quantum computing 
\cite{ahar,kempe,terhal}, and so development of adiabatic quantum algorithms
continues to be of great interest.

\begin{acknowledgments}
We thank W. G. Macready and D. Dahl for many interesting discussions, 
and F. G. thanks T. Howell III for continued support.
\end{acknowledgments}

\appendix

\section{Quantum adiabatic theorem}
\label{appendixQAT}

The question of how the state of a physical system changes when the system's 
environment undergoes a slow variation is an old one. For quantum systems, the 
answer is contained in the quantum adiabatic theorem which was proved by Born 
and Fock not long after the birth of quantum mechanics \cite{BF}. Subsequent 
work relaxed a number of the assumptions underlying the original proof
\cite{Kato,Avron1,Nenciu,Friedrichs,Hagedorn,Avron2,Avron3,Avron4,Narnhofer,
Martinez}, thereby widening the theorem's range of validity. As the physical 
setting for the quantum adiabatic theorem occurs often, it forms the foundation 
for many important applications in atomic, molecular, and chemical physics. 
Recently, it has been used as the basis for a novel alternative approach to 
quantum computing known as adiabatic quantum computing \cite{farhi,Farhi2}. 
In this Appendix we provide a brief review of the quantum adiabatic 
theorem (Appendix~\ref{appA1}) and then describe how it is used in adiabatic 
quantum computing (Appendix~\ref{appA2}). 

\subsection{Quantum adiabatic theorem}
\label{appA1}

Consider a quantum system coupled to an environment that changes slowly 
over a time $T$. The dynamical evolution of its state $|\psi (t)\rangle$ is 
determined by the Schrodinger equation, which will be driven by a 
slowly-varying Hamiltonian $H(t)$. The system's Hilbert space is assumed to be 
finite-dimensional with dimension $d$. Thus at each instant $t$ there will be 
$d$ instantaneous energy eigenstates $|E_{k}(t)\rangle$ satisfying
\begin{equation}
H(t) |E_{k}(t)\rangle = E_{k}(t)|E_{k}(t)\rangle ,
\label{insteig}
\end{equation} 
with $E_{0}(t)\leq E_{1}(t)\leq \cdots \leq E_{d-1}(t)$. 

Now suppose that we initially prepare the quantum system in the instantaneous 
energy eigenstate $|E_{k}(0)\rangle$ of the initial Hamiltonian $H(0)$. The 
quantum adiabatic theorem states that, in the limit $T\rightarrow\infty$, the 
final state $|\psi (T)\rangle$ will be (to within a phase factor) the 
instantaneous energy eigenstate $|E_{k}(T)\rangle$ of the final Hamiltonian 
$H(T)$. Note that in our discussions of adiabatic quantum computing, the 
initial state will always be the ground state of $H(0)$: $|\psi (0)\rangle = 
|E_{0}(0)\rangle$. 

For reasons that will become clear below, the energy gap, $\Delta (t) = E_{1}
(t)-E_{0}(t)$, separating the two lowest instantaneous energy-levels proves to 
be extremely important in adiabatic quantum computing. Although it is possible 
to prove the quantum adiabatic theorem for systems with a vanishing energy gap 
\cite{Avron4}, the rate of convergence to the adiabatic limit can be 
arbitrarily slow. On the other hand, when the gap $\Delta (t)$ is 
non-vanishing for $0\leq t \leq T$, it is possible to estimate how large $T$ 
must be for the dynamics to be effectively adiabatic. Starting from the 
observation that for adiabatic dynamics there will be negligible probability to
find the quantum system at $t=T$ in an energy-level other than the ground state,
a straightforward analysis \cite{messiah,Schiff} leads to the following 
adiabaticity constraint:  
\begin{equation}
T\gg \frac{\hbar M}{\Delta^{2}} ,
\label{adcon}
\end{equation}
where
\begin{eqnarray}
M & = & \max_{0\leq s\leq 1}\left| \langle E_{1}(s)|\frac{d\tilde{H}(s)}{ds}
|E_{0}(s)\rangle \right| , \label{newMdef}\\
\Delta & = & \min_{0\leq s\leq 1}\left[ E_{1}(s) - E_{0}(s)\right] ,
\label{mingapdef}
\end{eqnarray}
$s=t/T$, and $\tilde{H}(s) = H(sT)$. The quantity $\Delta$ is the minimum 
energy gap arising during the adiabatic evolution. 

The quantum adiabatic theorem provides a mechanism for traversing a path 
$|\psi (t)\rangle$ through Hilbert space that begins at a given state 
$|\psi_{i}\rangle$ and ends at a desired final state $|\psi_{f}\rangle$. To 
see this, let $H_{i}$ and $H_{f}$ be local Hermitian operators whose ground 
states are $|\psi_{i}\rangle$ and $|\psi_{f}\rangle$, respectively. An 
Hermitian operator is local if it couples at most $k$ particles, with $k$ 
finite. Suppose that we can apply a time-dependent Hamiltonian $H(t)$ over a 
time-interval $0\leq t\leq T$, with $H(0)=H_{i}$, and $H(T)=H_{f}$. Suppose 
further that we prepare our quantum system in the ground state $|\psi_{i}
\rangle$ of $H_{i}$, and then apply $H(t)$ to it. In the limit $T\rightarrow
\infty$, the quantum adiabatic theorem guarantees that the system at time $T$
will be in the desired state $|\psi_{f}\rangle$. The outcome is thus a 
continuous path from $|\psi_{i}\rangle$ to $|\psi_{f}\rangle$.

\subsection{Adiabatic quantum computing}
\label{appA2}

To connect this discussion to quantum computing, imagine that there is a 
computational problem we would like to solve, and that we are 
able to construct a local Hamiltonian $H_{f}$ whose ground state $|\psi_{f}
\rangle$ encodes the solution to our problem. Often, the computational basis 
states can be chosen to be the eigenstates of $H_{P}$. Our GI problem 
Hamiltonian $H_{P}$ is an example of such a Hamiltonian (see 
Section~\ref{sec3}). Let $H_{i}$ be a local Hamiltonian operator whose 
ground state $|\psi_{i}\rangle$ is easy to prepare 
(e.~g.~see Eq.~(\ref{HIdef})). In adiabatic quantum computing, the procedure 
presented in the preceding paragraph is applied with $|\psi (0)\rangle = 
|\psi_{i}\rangle$ and $H(t)$ tracing out a path from $H_{i}$ to $H_{f}$ (in 
the space of Hermitian operators). Originally, Ref.~\cite{farhi}
chose $H(t)$ to linearly interpolate from $H_{i}$ to $H_{f}$:
\begin{equation}
H(t) = \left( 1 - t/T\right) H_{i} + \left( t/T\right) H_{f}.
\label{Htdef}
\end{equation}
Writing $s=t/T$ and $\tilde{H}(s) = H(sT)$ gives
\begin{equation}
\tilde{H}(s) = \left( 1 - s\right) H_{i} + s H_{f} .
\label{tildeH}
\end{equation}
More general interpolation schemes are possible: $H(t) = A(t)H_{i} + B(t)H_{f}$,
where we require $A(0) = 1$ ($B(0) = 0$) and $A(T) = 0$ ($B(T) = 1$). See,
for example, Refs.~\cite{Rol&Cerf,Gaitan3,bian}. By 
choosing $T$ sufficiently large, the final state $|\psi (T) \rangle$ can be 
brought arbitrarily close to $|\psi_{f}\rangle$. An appropriate measurement 
then yields $|\psi_{f} \rangle$ with probability close to $1$, and thus yields 
the desired solution to our computational problem. Adiabatic quantum computing 
thus finds the solution to a computational problem by homing in on the 
ground state $|\psi_{f}\rangle$ of $H_{f}$ which encodes the solution. The 
homing mechanism is provided by the quantum adiabatic theorem. It has been
shown that adiabatic quantum computing has the same computational power as 
the circuit model for quantum computing \cite{ahar,kempe,terhal}. It thus 
provides an important 
alternative approach to quantum computing that is especially well suited to 
problems that reduce to quantum state generation. 

We are now in a position to state the protocol for the adiabatic quantum 
evolution (AQE) algorithm \cite{farhi}:
\begin{enumerate}
\item Prepare an $n$-qubit quantum register in the ground state $|\psi_{i}
\rangle$ of $H_{i}$.
\item At $t=0$, apply $H(t)$ to the quantum register for a time $T$.
\item At time $t=T$, measure the qubits in the computational basis.
\end{enumerate}
Because the computational basis states are typically the eigenstates of 
$H_{f}$, the final measurement leaves the qubits in an eigenstate of $H_{f}$. 
In the adiabatic limit $T\rightarrow\infty$, the final measurement leaves the 
qubits in the ground-state of $H_{f}$ (which encodes the solution we are 
trying to find) with probability $P_{success}\rightarrow 1$.

Now for large, but finite $T$, the Schrodinger dynamics is approximately 
adiabatic. Thus, with probability $1-\epsilon$, the measurement returns
the problem solution. In this case the AQE algorithm must be run more than 
once. Suppose we run it $\kappa$ times. The probability that we do 
\textit{not\/} get the problem solution in \textit{any\/} of the $\kappa$ runs 
is $\epsilon^{\kappa}$. We can make this probability take 
an arbitrarily small value $\tilde{\epsilon}$ by choosing $\kappa\sim 
\mathcal{O}(\log (1/\tilde{\epsilon}))$. Thus with probability arbitrarily 
close to $1$, one of the $\kappa$ measurement results will yield the problem
solution.

The adiabaticity constraint (see Eq.~(\ref{adcon})) specifies a lower bound 
which the runtime $T$ must exceed if the Schrodinger dynamics is to be 
effectively adiabatic. In all applications of interest to date, the matrix 
element $M$ appearing in this constraint scales polynomially with problem
size $N$. So long as this is true, Eq.~(\ref{adcon}) indicates that the scaling
behavior of the runtime $T(N)$ is determined by the scaling behavior of the 
minimum gap $\Delta (N)$. Now, if at $t=0$ the quantum register is prepared 
in the ground state of the initial Hamiltonian $H(0)$, and its dynamics is 
effectively adiabatic, its state at later times will be effectively restricted 
to the subspace spanned by the instantaneous ground and first excited states. 
Standard arguments \cite{llQM} indicate that, in the absence of symmetry, 
these two energy-levels will typically not cross. Thus the minimum gap 
$\Delta (N)$ will typically not vanish, and Eq.~(\ref{adcon}) indicates that 
an effectively adiabatic dynamics can be obtained with finite $T(N)$. An
algorithm, classical or quantum, is said to efficiently (inefficiently) solve
a computational problem if its runtime scales polynomially (super-polynomially)
with problem size. Thus, if $\Delta (N)$ scales inverse polynomially 
(super-polynomially) with $N$, then $T(N)$ will scale polynomially
(super-polynomially) with $N$, and the AQE algorithm will be an efficient
(inefficient) algorithm. We see that the scaling behavior of the minimum gap 
$\Delta (N)$ largely controls the computational complexity of the AQE 
algorithm. \\

\section{Embedding Procedure for D-Wave Hardware}
\label{appendixEmBed}

\begin{figure}
\includegraphics[width=7cm]{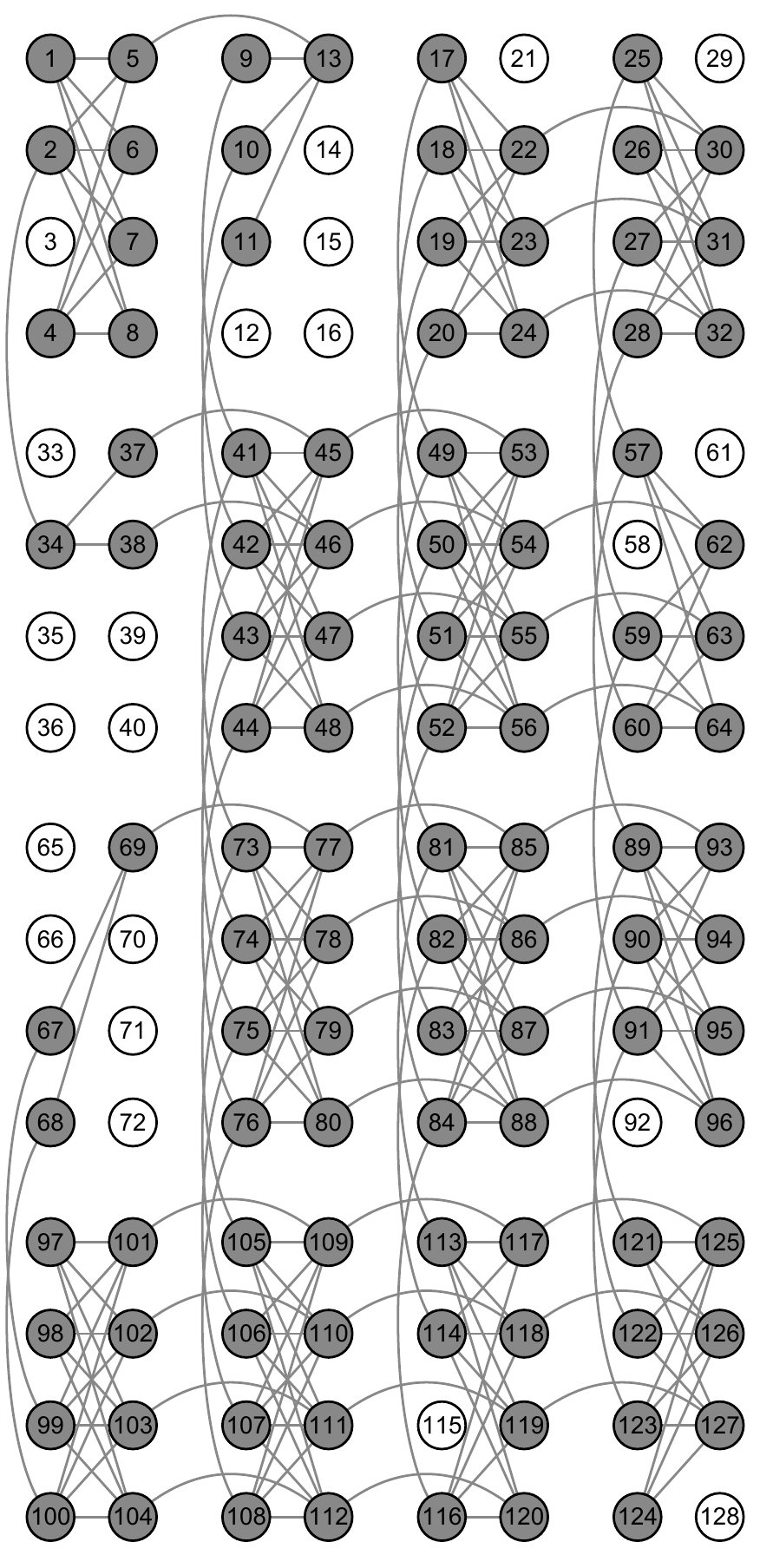}
\caption{\textbf{Layout of qubits and couplers for a D-Wave One processor.}
The processor architecture is a $4\times 4$ array of unit cells, with each unit
cell containing 8 qubits. Within a unit cell, each of the $4$ qubits in the 
left-hand partition (LHP) connects to all $4$ qubits in the right-hand 
partition (RHP), and vice versa. A qubit in the LHP (RHP) of a unit cell also 
connects to the corresponding qubit in the LHP (RHP) in the units cells above 
and below (to the left and right of) it. Most qubits couple to 6 neighbors. 
Qubits are labeled from 1 to 128, and edges between qubits indicate couplers 
which may take programmable values. Grey qubits indicate usable qubits, while 
white qubits indicate qubits which, due to fabrication defects, could not be
calibrated to operating tolerances and were not used.
} \label{qubitFig}
\end{figure}

This Appendix briefly describes the embedding procedure used in Ref.~\cite{bian}
to program a non-Ising problem Hamiltonian $H_{P}$ onto a D-Wave One processor.
The processor architecture is shown in Fig.~\ref{qubitFig}.
This procedure also applies to a D-Wave Two processor.

As discussed in Section~\ref{sec3}, the GI algorithm presented in this paper 
constructs $H_{P}$ to: (i)~be diagonal in the computational basis 
$\{ |a_{0}\cdots a_{L-1}\rangle : a_{i} = 0,1\}$; and 
(ii)~have eigenvalues $C(a)$, where $a = a_{0}\cdots a_{L-1}$, $L$ is the 
number of qubits, and $C(a)$ is given by 
Eqs.~(\ref{costfuncdefv2})-(\ref{C3def}) with $s\rightarrow a$. The cost 
function $C(a)$ is not yet ready for experimental implementation for two 
reasons. First, there are $k$-qubit interactions with $k>2$ which cannot be 
implemented as the processor can only couple pairs of qubits; 
and second, two-qubit couplings may not correspond to available couplings on 
the processor (see Fig.~\ref{qubitFig}). Procedures for removing each of 
these obstacles are presented, respectively, in Appendices~\ref{appendixB1} 
and \ref{appendixB2}. We summarize these procedures in 
Appendix~\ref{appendixB3}. To keep the discussion concrete, we examine a 
single $k$-qubit coupling term $A = a_{1}\cdots a_{k}$ with $a_{i} = 0,1$. 
The resulting procedure must then be applied to each term in $C(a)$.\\

\subsection{Reduction to pairwise coupling}
\label{appendixB1}

Here we describe how to reduce a $k$-qubit coupling term $A = a_{1}\cdots 
a_{k}$ with $a_{i} = 0,1$ to a sum of $2$-qubit (viz.~pairwise) coupling terms. 
We first show how to reduce a $3$-qubit coupling term to pairwise coupling 
terms (Appendix~\ref{appendixB11}), and then use lessons learned to reduce 
the $k$-qubit term $A$ to pairwise coupling (Appendix~\ref{appendixB12}).

\subsubsection{3-qubit case}
\label{appendixB11}

 We begin by showing how to reduce a $3$-qubit coupling term 
$a_{1}a_{2}a_{3}$ to pairwise coupling by introducing: (i)~an ancillary 
variable $b$ which takes only two values $\{0,1\}$; and (ii)~the penalty 
function
\begin{equation}
P(a_1,a_2;b) = a_1 a_2 - 2(a_1+a_2)b+3b . \label{penFuncEq}
\end{equation}
Notice that $P(a_{1},a_{2};b) = 0$ ($>0$) when the input values for $a_{1}$,
$a_{2}$, and $b$ satisfy $b=a_{1}a_{2}$ ($b\neq a_{1}a_{2}$). Now consider
the quadratic cost function 
\begin{displaymath}
h(b) = ba_{3}+ \mu P(a_{1}, a_{2};b)
\end{displaymath}
for given values of $\mu$, $a_{1}$, and $a_{2}$. For $\mu$ sufficiently large,
$h(b)$ is minimized when the value of $b$ satisfies the equality constraint 
$b^{\ast}=a_{1}a_{2}$. As noted above, for this optimal value of $b$, the 
penalty function $P(a_{1},a_{2};b^{\ast}) = 0$, and so the optimum cost
$h(b^{\ast})$ is
\begin{eqnarray*}
h(b^{\ast}) & = & b^{\ast}a_{3} + P(a_{1},a_{2};b^{\ast}) \nonumber\\
 & = & a_{1}a_{2}a_{3},\\
\end{eqnarray*}
where $b^{\ast} = a_{1}a_{2}$ has been used in going from the first to the
second line. Thus, for values of $b$ satisfying the equality constraint $b = 
a_{1}a_{2}$, the cost function $h(b)$, which is a sum of $2$-qubit coupling 
terms, reproduces the $3$-qubit coupling term $a_{1}a_{2}a_{3}$. By choosing 
$\mu$ sufficiently large, values of $b$ that do not satisfy the equality 
constraint can be pushed to large cost (viz.~energy), making such
$b$ values inaccessible during adiabatic quantum evolution.

\subsubsection{k-qubit case}
\label{appendixB12}

To reduce the $k$-qubit coupling term $A = a_{1}\cdots a_{k}$ to pairwise 
coupling we: (i)~introduce ancillary bit variables $b_{2},\cdots , b_{k-1}$; 
and (ii)~impose the constraints $b_{k-1}=a_{k-1}a_{k}$ and $b_{j}=a_{j}
b_{j+1}$ ($j=2,\cdots ,k-2$) through the penalty function 
\begin{displaymath}
P(\mathbf{a};\mathbf{b}) = P(a_{k-1},a_k;b_{k-1})+ \sum_{j=2}^{k-2}
P(a_j, b_{j+1};b_j),
\end{displaymath}
where $\mathbf{a} = (a_{1}, \cdots , a_{k})$, $\mathbf{b} = (b_{2}, \cdots ,
b_{k-1})$ and $P(a,b;c)$ is defined in Eq.~(\ref{penFuncEq}). The quadratic
cost function $C_{A}(\mathbf{a},\mathbf{b})$ is defined to be
\begin{displaymath}
C_{A}(\mathbf{a},\mathbf{b}) =  a_{1} b_{2}
                                 + \mu\, P(\mathbf{a};\mathbf{b}) .
\end{displaymath}
We require that the optimal values $(\mathbf{a}^{\ast},\mathbf{b}^{\ast})$ 
satisfy the $k-1$ imposed constraints so that $P(\mathbf{a}^{\ast},
\mathbf{b}^{\ast}) = 0$. For optimal values, the cost function evaluates to 
\begin{eqnarray*}
C_{A}(\mathbf{a}^{\ast},\mathbf{b}^{\ast}) & = & a_{1}^{\ast}b_{2}^{\ast}
             + \mu P(\mathbf{a}^{\ast},\mathbf{b}^{\ast})  \nonumber\\
 & = & a_{1}^{\ast}a_{2}^{\ast}b_{3}^{\ast} \nonumber\\
 & = & \vdots \nonumber\\
 & = & a_{1}^{\ast}\cdots a_{k-2}^{\ast}b_{k-1}^{\ast} \nonumber\\
 & = & a_{1}^{\ast}\cdot a_{k}^{\ast} ,
\end{eqnarray*}
where (in the interests of clarity) we have introduced the constraints one at a
time in going from one line to the next. Here $\mu$ is a penalty weight whose 
value is chosen large enough to freeze out non-optimal values of $\mathbf{a}$ 
and $\mathbf{b}$ during adiabatic quantum evolution. Thus, for values of
$\mathbf{a}$ and $\mathbf{b}$ satisfying the $k-1$ equality constraints, and
for $\mu$ sufficiently large, the cost function $C_{A}(\mathbf{a},\mathbf{b})$, 
which is a sum of $2$-qubit coupling terms, reproduces the $k$-qubit coupling 
term $A = a_{1}\cdots a_{k}$ as desired. 

\subsection{Matching required couplings to hardware couplings}
\label{appendixB2}

A cost function with only pairwise coupling such as results from the procedure 
described in Appendix~\ref{appendixB1} may still not be experimentally 
realizable on a D-Wave processor as the pairwise couplings arising in 
$C_{A}(\mathbf{a},\mathbf{b})$ may not match the two-qubit couplings 
available on the processor. The primal graph of a quadratic cost function 
such as $C_{A}(\mathbf{a},\mathbf{b})$ is the graph whose vertices are 
the qubit variables, and whose edges indicate pairwise-coupled qubits. An 
arbitrary primal graph can be embedded into a sufficiently large qubit graph 
having the structure of  Fig.~\ref{qubitFig}. An embedding maps a primal 
graph vertex to one or more vertices in the qubit graph, where the image 
vertices form a connected subgraph of the qubit graph. The string of connected 
qubits are linked together with strong ferromagnetic couplings so that in the 
lowest energy state, these qubits have identical Bloch vectors. For example, 
to couple qubits 104 and 75  in Fig.~\ref{qubitFig} (which are not directly 
coupled) with coupling strength $\mathcal{J}$, we ferromagnetically couple 
qubits 104, 112, and 107 using strongly negative $J_{104,112}$ and 
$J_{107,112}$ values. Qubits 107 and 75 are directly coupled by the 
processor and so the desired coupling $\mathcal{J}$ is applied to the 
edge (viz.~coupler $J_{107,75}$) connecting qubits 107 and 75: 
$J_{107,75} = \mathcal{J}$. The ferromagnetic chain thus effects the
desired coupling of qubits 104 and 75. This embedding procedure must be 
carried out for each pair of primal graph vertices joined by an edge whose 
associated qubits are not directly coupled in the processor architecture.

\subsection{Summary}
\label{appendixB3}

By combining the procedures described in this Appendix it is possible to 
transform any cost function $C(a)$ into a quadratic cost function with
pairwise couplings that matches the couplings specified by the processor
architecture. The trade-off is the introduction of ancilla qubits that are
needed to reduce the coupling interactions to pairwise coupling and to 
match the $2$-qubit couplings available on the processor.

\end{document}